\documentclass[final,5p,times,twocolumn,numbers,sort&compress]{elsarticle}
\usepackage{natbib}

 \usepackage{graphicx}
\usepackage{subfigure}
\usepackage{color}

\usepackage{latexsym,bm}

\usepackage{amssymb}
\usepackage{amsmath}

\usepackage[linesnumbered,ruled]{algorithm2e}
\usepackage{algorithmic}

\usepackage{cases}

\usepackage{threeparttable}
\usepackage{multirow}

\usepackage[colorlinks,linkcolor=red]{hyperref}

\newtheorem{Def}{Definition}
\newtheorem{thm}{Theorem}
\newtheorem{lem}{Lemma}

\newtheorem{pro}{Proposition}
\newtheorem{rem}{Remark}

\newcommand{\norm}[1]{\left\Vert#1\right\Vert}
\newcommand{\abs}[1]{\left| #1 \right|}
\newcommand{\pref}[1]{(\ref{#1})}

\allowdisplaybreaks

\begin{document}

\begin{frontmatter}

\title{Constructing Trivariate B-splines with Positive Jacobian by Pillow Operation and Geometric Iterative Fitting}

\author[add1,add2]{Hongwei Lin\corref{cor1}}
\author[add2]{Hao Huang}
\author[add1]{Chuanfeng Hu}
    \cortext[cor1]{Corresponding author: phone number: 86-571-87951860-8304, fax number: 86-571-88206681, email:
    hwlin@zju.edu.cn}
\address[add1]{School of Mathematics, Zhejiang University, Hangzhou, 310027, China}
\address[add2]{State Key Lab. of CAD\&CG, Zhejiang University, Hangzhou, 310027, China}

\date{}

\begin{abstract}
  The advent of isogeometric analysis has prompted a need for methods to generate Trivariate B-spline Solids (TBS) with positive Jacobian. However, it is difficult to guarantee a positive Jacobian of a TBS since the geometric pre-condition for ensuring the positive Jacobian is very complicated. In this paper, we propose a method for generating TBSs with guaranteed positive Jacobian.
  For the study, we used a tetrahedral (tet) mesh model and segmented it into sub-volumes using the pillow operation. Then, to reduce the difficulty in ensuring a positive Jacobian, we separately fitted the boundary curves and surfaces and the sub-volumes using a geometric iterative fitting algorithm .
  Finally, the smoothness between adjacent TBSs is improved. The experimental examples presented in this paper demonstrate the effectiveness and efficiency of the developed algorithm.
\end{abstract}

\end{frontmatter}

%%%%%%%%%%%%%%%%%%%%%%%%%%%%%%%%%%%%%%%%%%%%%%%%%%%%%%%%%%%%%%%%%%%%

%-----------------------------------------------------------------------------------------
% Section: Introduction
%-----------------------------------------------------------------------------------------

 \section{Introduction}
 \label{sec:introduction}

 While the traditional geometric design community focuses on the design of curves and surfaces, the advent of Isogeometric Analysis (IGA)~\cite{hughes2005isogeometric} has made the development of methods for designing Trivariate B-spline Solids (TBSs) imperative.
 In IGA, a \emph{valid} TBS should have a positive Jacobian value at every point in its domain. A negative Jacobian value at any points in the domain of the TBS can render the IGA invalid.

 As is widely recognized, the generation of a valid TBS is difficult, owing to two key factors:

 The geometric condition for ensuring that a TBS is valid, i.e., the Jacobian value at every point is positive, is highly nonlinear. Hence, it is theoretically very difficult to guarantee the \emph{validity} of a TBS. The state-of-the-art methods for generating TBS usually transform the validity problem into an optimization problem. However, the optimization is prone to fail owing to the high nonlinearity of the objective function. Therefore, to generate valid TBSs, the high nonlinearity of the objective function should be reduced and an efficient method should be developed for solving the optimization problem.

 The regions with negative Jacobian are usually concentrated near the boundary curves where adjacent surfaces are smoothly stitched (refer to Fig.~\ref{fig:negative_jac}). To reduce negative Jacobian values, the input model should be segmented along the boundary curves using the \emph{pillow} operation~\cite{mitchell1995pillowing}, a method originally developed for improving the quality of hexahedral meshes.

 In this paper, we developed a method that can generate a TBS with a guaranteed positive Jacobian. Using a tetrahedral (tet) mesh with six surfaces segmented on its boundary mesh as the input, we first partition the tet mesh model into seven sub-volumes using the \emph{pillow} operation.
 After each of them is parameterized into a cubic parameter domain, seven initial valid TBSs are constructed. Moreover, starting with the initial valid TBSs, the boundary curves, boundary surfaces and the TBSs are fitted by a geometric iterative fitting algorithm, known as the Geometric Feasible Direction (GFD) algorithm.
 In each iteration of the GFD algorithm, the movements of the control points are restricted inside a feasible region to ensure the validity. Finally, the smoothness between two adjacent TBSs is improved by the GFD algorithm. In this way, the validity of the generated TBSs is guaranteed with desirable smoothness between adjacent TBSs.

 The structure of this paper is as follows:
 Section~\ref{subsec:related_work}
 provides an overview of the additional work performed to develop this study.
 Section~\ref{sec:overview} presents an overview of the proposed method.
 In Section~\ref{sec:conditions},
    we introduce the validity conditions for B-spline curves, surfaces, and TBSs, and the geometric continuity between TBSs.
 In Section~\ref{sec:gfd}, the Geometric Feasible Direction (GFD) algorithm
    is developed.
 After elucidating the details of the developed method in
    Section~\ref{sec:algorithm},
    the experimental results are illustrated in Section~\ref{sec:result}.
 Lastly, Section~\ref{sec:conclusion} concludes the paper.

 %----------------------------------------------------------------------------
 % Subsection:
 %----------------------------------------------------------------------------
 \subsection{Related work}
 \label{subsec:related_work}

 In this section, work related to TBS modeling and geometric iterative fitting is briefly reviewed.

 \textbf{TBS modeling:}
 To analyze the arterial blood flow by IGA, a trivariate NURBS-solid modeling the artery was constructed using a skeleton-based method~\citep{zhang2007patient}. Following volume parameterization by a harmonic function, a cylinder-like trivariate B-spline solid was generated with a singular centric curve~\cite{martin2009volumetric}. Moreover, trivariate B-spline solids with positive Jacobian values were produced from boundary representations using optimization-based approaches~\citep{xu2013analysis, wang2014optimization}. However, the optimization method may fail when the objective function is highly nonlinear. Based on the given boundary conditions and guiding curves, a NURBS solid was constructed to model a swept volume by the variational approach~\citep{aigner2009swept}.

 Moreover, trivariate T-spline solids are employed in IGA owing to their adaptive refinement capability~\cite{sederberg2004t}. To fill a genus-zero triangular mesh, a T-spline solid was constructed after mesh untangling and smoothing~\citep{escobar2011new}. Starting from a boundary surface triangulation with genus-zero, Zhang et al. developed a mapping-based method to generate rational trivariate solid T-splines~\citep{zhang2012solid}.
 Furthermore, to fill a boundary triangulation with arbitrary genus, a polycube mapping-based algorithm for constructing T-spline solids was proposed in~\citep{wang2013trivariate}. In~\citep{liu2014volumetric}, a volumetric T-spline was constructed for filling a B-rep model using Boolean operations, polycube mapping, and octree subdivision.
 The inputs for the aforementioned methods are triangular meshes. When the input was a genus-zero T-spline surface model, Zhang et al. constructed a solid T-spline whose boundary exactly conformed to the given T-spline surface model~\citep{zhang2013conformal}.

 On the other hand, Catmull-Clark subdivision solids were developed to model the computational domain in IGA~\citep{burkhart2010iso}. Additionally, other representations for spline solids include simplex splines~\citep{hua2004multiresolution} and polycube splines~\citep{li2010generalized,wang2012restricted}.

 The methods described above usually generate a trivariate solid to fill a given B-rep model. However, the generation of a TBS by fitting a tet mesh model is much easier than by filling a B-rep model because it is very easy to produce a tet mesh using popular software such as, TetGen~\cite{sitetgen}, and NetGen~\cite{schoberl1997netgen}. Hence, it is feasible to generate a TBS by fitting a tet mesh model.
 In Ref.~\citep{lin2015constructing}, a tet mesh model is fitted by the geometric iterative method to generate a TBS. However, there are some regions with negative Jacobian values close to the boundary.
 In this experiment, a tet mesh is first segmented into seven sub-volumes, each of which is fitted with a TBS by a geometric iterative fitting method. In this way, the generated TBSs are ensured to be valid, i.e. the Jacobian value at any point of the TBSs is positive.

 In conclusion, the representations of a trivariate solid include B-spline solid, NURBS solid, T-spline solid, subdivision solid, etc. The input for a solid generation method is either a boundary representation, such as B-spline surface, a triangular mesh, or a volume mesh, such as a tet mesh. In our method, the input model is segmented to generate the valid TBS. While a tet meshes can be partitioned into sub-volumes, the boundary representations, as solids, are hard to be segmented owing to the loss of inner information. Therefore, tet meshes are taken as inputs for our method.

 %%%%%%%%%%%%%%%%%%%%%%%%%%%%%%%%%%%%%%%%%%%%%%%%%%%%%%%%%%%%%%%%%%%%%%%%%%%%%%%%%%

 \textbf{Geometric iterative fitting:}
 Geometric Iterative Fitting (GIF), also called Progressive-Iterative Approximation (PIA), was first developed in~\citep{lin2004constructing,lin2005totally}. The GIF method endows iterative methods with geometric meanings, facilitating the handling of geometric problems appearing in the field of geometric design.
 It was proved that the GIF method is convergent for B-spline fitting~\citep{lin2011extended,deng2014progressive}, NURBS fitting~\citep{shi06iterative}, T-spline fitting~\citep{lin2013efficient}, subdivision surface fitting~\citep{cheng2009loop, fan2008subdivision, chen2008progressive}, as well as curve and surface fitting with totally positive basis~\citep{lin2005totally}. Moreover, the GIF methods have been employed in some applications such as reverse engineering~\citep{kineri2012b,yoshihara2012topologically}, curve design~\citep{okaniwa2012uniform}, surface-surface intersection~\citep{lin2014affine}, etc.
 In Ref.~\citep{lin2015constructing}, the GIF method was employed to fit a tet mesh model with a TBS, showing that the GIF method is capable of solving singular linear systems. In this paper, the GIF method is used to fit several sub-volumes with valid TBSs.

%----------------------------------------------------------------------------
 % Section:
 %----------------------------------------------------------------------------
 \section{Algorithm overview}
 \label{sec:overview}

%------------------------------------------------------------------------------
%-----------------------------Figure-------------------------------------------
\begin{figure*}[!htb]
  \centering
    \includegraphics[width=0.7\textwidth]{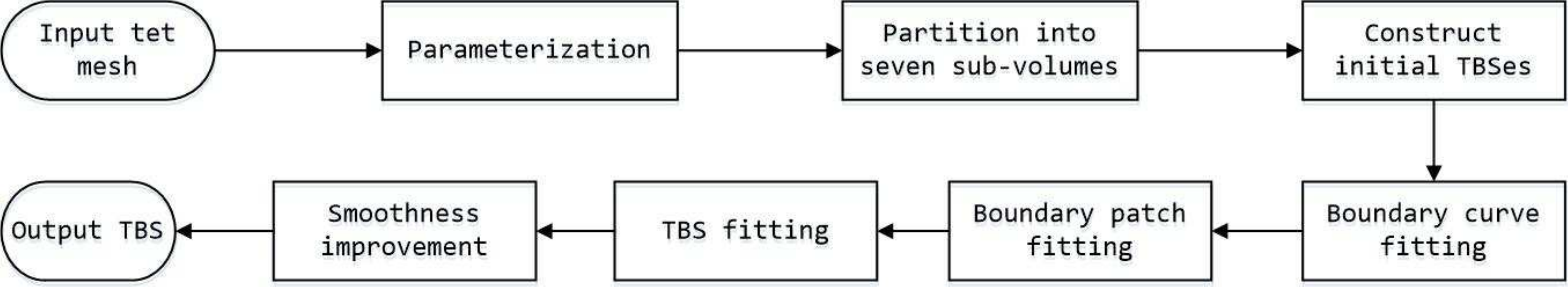}
      \caption{Overview of the TBS generation algorithm developed in this paper.}
   \label{fig:overview}
\end{figure*}
%------------------------------------------------------------------------------
 The TBS generation algorithm is demonstrated in Fig.~\ref{fig:overview}.
 The input to this algorithm is a tet mesh with six boundary surfaces (Fig.~\ref{subfig:input}).
 The tet mesh is first parameterized into a cubic parametric domain. Then, the cubic domain is segmented into seven sub-domains, as shown in the (Fig.~\ref{subfig:para_domain}).
 By mapping the sub-domains into the input tet mesh, it is divided into seven sub-volumes. After the initial TBSs are constructed, the boundary curves, boundary surfaces, and the seven TBSs are respectively fitted using the GFD algorithm. Finally, the smoothness between adjacent TBSs is improved. In subsequent sections, the details of the TBS generation algorithm will be elucidated.

%-----------------------------------------------------------------------------
% Section:
%-----------------------------------------------------------------------------
 \section{Validity conditions and geometric continuity between TBSs}
 \label{sec:conditions}

 In this paper, what we want to generate is a composition of valid TBSs with
    as desirable as possible smoothness between adjacent TBSs.
 So, the validity conditions and geometric continuity definition
    between TBSs should be clarified.

%-----------------------------------------------------------------------------
% Section:
%-----------------------------------------------------------------------------
 \subsection{Validity conditions}
 \label{subsec:validity_condition}

 In this section, the validity conditions for a B-spline curve,
    surface, and TBSs are presented, respectively.

 %--------------------------------------------------------------------
 % Definition
 %--------------------------------------------------------------------
 \begin{Def}
 \begin{enumerate}
   \item[1)] A B-spline curve is valid, if its derivative vector is nonzero at any parameter in its domain;
   \item[2)] A B-spline surface is valid, if its normal vector is nonzero at any point in its parametric domain;
   \item[3)] A TBS $\bm{H}(u,v,w)$ is valid, if $\bm{H}_u \cdot (\bm{H}_v \times \bm{H}_w) \neq 0$ at any point in its parametric domain.
 \end{enumerate}
 \end{Def}

 Given a B-spline curve with control point $\bm{P}_i, i = 0,1,\cdots,m$,
    and denoting the difference vectors as,
    $$\bm{T}_i = \frac{\bm{P}_{i+1}-\bm{P}_i}{\norm{\bm{P}_{i+1}-\bm{P}_i}},\ i=0,1,\cdots,m-1,$$
    the valid condition for the B-spline curve is,
 %----------------------------------------------------------------------
 % Proposition
 %----------------------------------------------------------------------
 \begin{pro}[Validity condition for B-spline curves] \label{pro:analysis_curve}
    A B-spline curve of degree $d$ is valid if the apertures of the minimum circular cones enclosing the difference vectors $\{\bm{T}_i, \bm{T}_{i+1}, \cdots, \bm{T}_{i+d-1} \}, \ i = 0,1,\cdots,m-d$, respectively,
    are all less than $\frac{\pi}{2}$.
 \end{pro}
 %----------------------------------------------------------------------

 The proof of Proposition~\ref{pro:analysis_curve} is evident.

 %\textbf{Proof:} If the condition of the proposition is true,
%    the derivative vector of the B-spline curve is nonzero at any parameter in its domain.
%    Therefore, the given B-spline curve is valid.
%    $\Box$

 \vspace{0.3cm}

 Moreover, suppose we are given a B-spline surface of degree $d_u \times d_v$
    with control points
    $$\bm{S}_{ij}, i=0,1,\cdots,m,\ j = 0,1,\cdots,n,$$
    and denote the difference vectors as
 \begin{equation}\label{eq:surface_diff_u}
   \bm{T}_{ij}^u = \frac{\bm{S}_{i+1,j}-\bm{S}_{ij}}{\norm{\bm{S}_{i+1,j}-\bm{S}_{ij}}},\ i = 0,1,\cdots,m-1, j=0,1,\cdots,n,
 \end{equation}
 and,
 \begin{equation}\label{eq:surface_diff_v}
   \bm{T}_{ij}^v = \frac{\bm{S}_{i,j+1}-\bm{S}_{ij}}{\norm{\bm{S}_{i,j+1}-\bm{S}_{ij}}},\ i = 0,1,\cdots,m, j = 0,1,\cdots,n-1.
 \end{equation}
 Let $\mathcal{M}_{IJ}, I = 0,1,\cdots,m-d_u, J=0,1,\cdots,n-d_v$ be the
    sub-control-polygon constituted by the control points,
 \begin{equation} \label{eq:sub_control_polygon}
      \begin{array}{cccc}
        \bm{S}_{IJ}   & \bm{S}_{I,J+1} & \cdots & \bm{S}_{I,J+d_v} \\
        \bm{S}_{I+1,J} & \bm{S}_{I+1,J+1} & \cdots  & \bm{S}_{I+1,J+d_v} \\
        \vdots & \vdots &   & \vdots \\
        \bm{S}_{I+d_u,J} & \bm{S}_{I+d_u,J+1} & \cdots & \bm{S}_{I+d_u,J+d_v} \\
      \end{array}
 \end{equation}
 Moreover, suppose $\bm{U}_{IJ}$ and $\bm{V}_{IJ}$ are the unit axis
    vectors of the minimum circular cones $\mathcal{C}^u_{IJ}$ and $\mathcal{C}^v_{IJ}$ enclosing the difference vectors $\bm{T}^u_{ij}$~\pref{eq:surface_diff_u} and $\bm{T}^v_{kl}$~\pref{eq:surface_diff_v} of the sub-control-polygon $\mathcal{M}_{IJ}$,
    starting from the apexes of cones, respectively.
 Then, a sufficient condition for the validity of a B-spline surface is
    presented as follows:
 %----------------------------------------------------------------------
 % Proposition
 %----------------------------------------------------------------------
 \begin{pro}[Validity condition for B-spline surfaces] \label{pro:analysis_surface}
    If $\bm{T}^u_{ij} \cdot \bm{U}_{IJ} > \bm{T}^u_{ij} \cdot \bm{V}_{IJ} \geq 0,\
        \text{and},\
      \bm{T}^v_{kl} \cdot \bm{V}_{IJ} > \bm{T}^v_{kl} \cdot \bm{U}_{IJ} \geq 0$,
      where $\bm{T}^u_{ij}$ and $\bm{T}^v_{kl}$ are defined on each sub-control-polygon $\mathcal{M}_{IJ},\ I = 0,1,\cdots,m-d_u, J=0,1,\cdots,n-d_v$,
      the B-spline surface is valid.
 \end{pro}

 \textbf{Proof:} Because,
    \begin{equation*}
      \bm{T}^u_{ij} \cdot \bm{U}_{IJ} > \bm{T}^u_{ij} \cdot \bm{V}_{IJ} \geq 0,\ \text{and},\
      \bm{T}^v_{kl} \cdot \bm{V}_{IJ} > \bm{T}^v_{kl} \cdot \bm{U}_{IJ} \geq 0,
    \end{equation*}
    we have,
    \begin{equation*}
      (\bm{T}^u_{ij} \times \bm{T}^v_{kl}) \cdot (\bm{U}_{IJ} \times \bm{V}_{IJ}) =
      (\bm{T}^u_{ij} \cdot \bm{U}_{IJ}) (\bm{T}^v_{kl} \cdot \bm{V}_{IJ}) -
      (\bm{T}^u_{ij} \cdot \bm{V}_{IJ}) (\bm{T}^v_{kl} \cdot \bm{U}_{IJ}) > 0.
    \end{equation*}
    It means that, all of the vectors $\bm{T}^u_{ij} \times \bm{T}^v_{kl}$ of
        the sub-control-polygon $\mathcal{M}_{IJ}$
        are on the same side of the plane with $\bm{U}_{IJ} \times \bm{V}_{IJ}$ as the normal vector.
    It makes the normal vector of the given B-spline surface nonzero
        at any point in its domain.
    So, the given B-spline surface is valid. $\Box$

 %-----------------------------------------------------------
 % Remark:
 %-----------------------------------------------------------
\begin{rem}
    Note that, the condition
    $\bm{T}^u_{ij} \cdot \bm{U}_{IJ} > \bm{T}^u_{ij} \cdot \bm{V}_{IJ} \geq 0,\
        \text{and},\
      \bm{T}^v_{kl} \cdot \bm{V}_{IJ} > \bm{T}^v_{kl} \cdot \bm{U}_{IJ} \geq 0$,
    means that the two minimum circular cones $\mathcal{C}^u_{IJ}$ and
    $\mathcal{C}^v_{IJ}$ are separate.
    According to the definition of normal vector,
    the closer to orthogonality the two unit axis vectors $\bm{U}_{IJ}$ and $\bm{V}_{IJ}$ of the circular cones $\mathcal{C}^u_{IJ}$ and $\mathcal{C}^v_{IJ}$,
    the better the validity of the B-spline surface.
\end{rem}

 Similarly, we can develop a sufficient condition for determining the
    validity of a TBS $\bm{H}(u,v,w)$ of degree $d_u \times d_v \times d_w$,
    $$\bm{H}(u,v,w) = \sum_i \sum_j \sum_k \bm{H}_{ijk} B_i^{d_u}(u)B_j^{d_v}(v)B_k^{d_w}(w),$$
    where $B_i^{d_u}(u), B_j^{d_v}(v),\ \text{and}\ B_k^{d_w}(w)$ are B-spline basis functions of degrees $d_u, d_v,\ \text{and}\ d_w$, respectively, and,
    $$\bm{H}_{ijk}, i=0,1,\cdots,m, j=0,1,\cdots,n, k = 0,1,\cdots,l,$$
    are control points.
 Denote the difference vectors as
 \begin{equation} \label{eq:diff_vec_tbs}
   \begin{split}
   \bm{T}^u_{ijk} & = \frac{\bm{H}_{i+1,j,k} - \bm{H}_{ijk}}
                         {\norm{\bm{H}_{i+1,j,k} - \bm{H}_{ijk}}},
   \bm{T}^v_{ijk}  = \frac{\bm{H}_{i,j+1,k} - \bm{H}_{ijk}}
                         {\norm{\bm{H}_{i,j+1,k} - \bm{H}_{ijk}}},\\
   \bm{T}^w_{ijk} & = \frac{\bm{H}_{i,j,k+1} - \bm{H}_{ijk}}
                         {\norm{\bm{H}_{i,j,k+1} - \bm{H}_{ijk}}}.
   \end{split}
 \end{equation}
 Moreover, letting
 \begin{equation} \label{eq:sub_grid}
 \begin{split}
    \mathcal{G}_{IJK},\ & I = 0,1,\cdots,m-d_u, J=0,1,\cdots,n-d_v,\\
                       & K = 0,1,\cdots,l-d_w,
 \end{split}
 \end{equation}
 be the sub-grid constituted by the control points
 \begin{equation*}
 \begin{split}
    \bm{H}_{ijk},\ & i=I,I+1,\cdots,I+d_u, j=J,J+1,\cdots,J+d_v, \\
                  & k=K,K+1,\cdots,K+d_w,
 \end{split}
 \end{equation*}
 we have,
 %If the three minimum circular cones $C_u, C_v$, and $C_w$,
%    enclosing $\bm{T}^u_{ijk}, \bm{T}^v_{ijk}$, and $\bm{T}^w_{ijk}$, respectively, are separated,
%    the TBS is valid.
% The closer to orthogonality the three circular cones $C_u, C_v$, and $C_w$,
%    the better the validity of the TBS.
 %----------------------------------------------------------------------
 % Proposition
 %----------------------------------------------------------------------
 \begin{pro}[Validity condition for TBSs] \label{pro:analysis_tbs}
    If $$\bm{T}^u_{i_u j_u k_u} \cdot (\bm{T}^v_{i_v j_v k_v} \times \bm{T}^w_{i_w j_w k_w}) > 0,$$
    where $\bm{T}^u_{i_u j_u k_u}, \bm{T}^v_{i_v j_v k_v}$ and
    $\bm{T}^w_{i_w j_w k_w}$ are defined on each sub-grid $\mathcal{G}_{IJK}$,
    the TBS $\bm{H}(u,v,w)$ is valid.
 \end{pro}

 \textbf{Proof:} The Jacobian value of the TBS $\bm{H}(u,v,w)$ at $(u,v,w)$
    is,
    \begin{equation*}
      \begin{split}
     & J(u,v,w)  = \bm{H}_u(u,v,w) \cdot (\bm{H}_v(u,v,w) \times
                \bm{H}_w(u,v,w)) \\
               & = \sum_{\mathcal{I}_u} \sum_{\mathcal{I}_v} \sum_{\mathcal{I}_w}
                    \alpha_{\mathcal{I}_u \mathcal{I}_v \mathcal{I}_w}
                   \left[\bm{T}_{i_u j_u k_u}^u \cdot (\bm{T}_{i_v j_v k_v}^v \times \bm{T}_{i_w j_w k_w}^w)\right]
                   B_{\mathcal{I}_u}(u) B_{\mathcal{I}_v}(v) B_{\mathcal{I}_w}(w),  \\
      \end{split}
    \end{equation*}
    where $\mathcal{I}_u = (i_u, i_v, i_w), \mathcal{I}_v = (j_u, j_v, j_w),
           \mathcal{I}_w = (k_u, k_v, k_w)$ are index sets, $\alpha_{\mathcal{I}_u \mathcal{I}_v \mathcal{I}_w} > 0$,
           and,
    \begin{equation*}
    \begin{split}
           & B_{\mathcal{I}_u}(u) = B_{i_u}^{d_u-1}(u) B_{i_v}^{d_u}(u) B_{i_w}^{d_u}(u),
           B_{\mathcal{I}_v}(v) = B_{j_u}^{d_v}(v) B_{j_v}^{d_v-1}(v) B_{j_w}^{d_v}(w),\\
           & B_{\mathcal{I}_w}(w) = B_{k_u}^{d_w}(w) B_{k_v}^{d_w}(v) B_{k_w}^{d_w-1}(w).
    \end{split}
    \end{equation*}
    Moreover, because the local support property of TBS,
        the Jacobian value $J(u,v,w)$ is determined by one of the
        sub-grid $\mathcal{G}_{IJK}$~\pref{eq:sub_grid}.
    Therefore, if $\bm{T}^u_{i_u j_u k_u} \cdot (\bm{T}^v_{i_v j_v k_v} \times \bm{T}^w_{i_w j_w k_w}) > 0$,
    where $\bm{T}^u_{i_u j_u k_u}, \bm{T}^v_{i_v j_v k_v}$ and
    $\bm{T}^w_{i_w j_w k_w}$ are defined on the sub-grid $\mathcal{G}_{IJK}$,
    it follows that
    $$\bm{H}_u(u,v,w) \cdot (\bm{H}_v(u,v,w) \times \bm{H}_w(u,v,w)) > 0,$$
    meaning that, the TBS $\bm{H}(u,v,w)$ is valid. $\Box$

 %------------------------------------------------------------------------------
%-----------------------------Figure-------------------------------------------
\begin{figure}[!htb]
  \centering
    \includegraphics[width=0.25\textwidth]{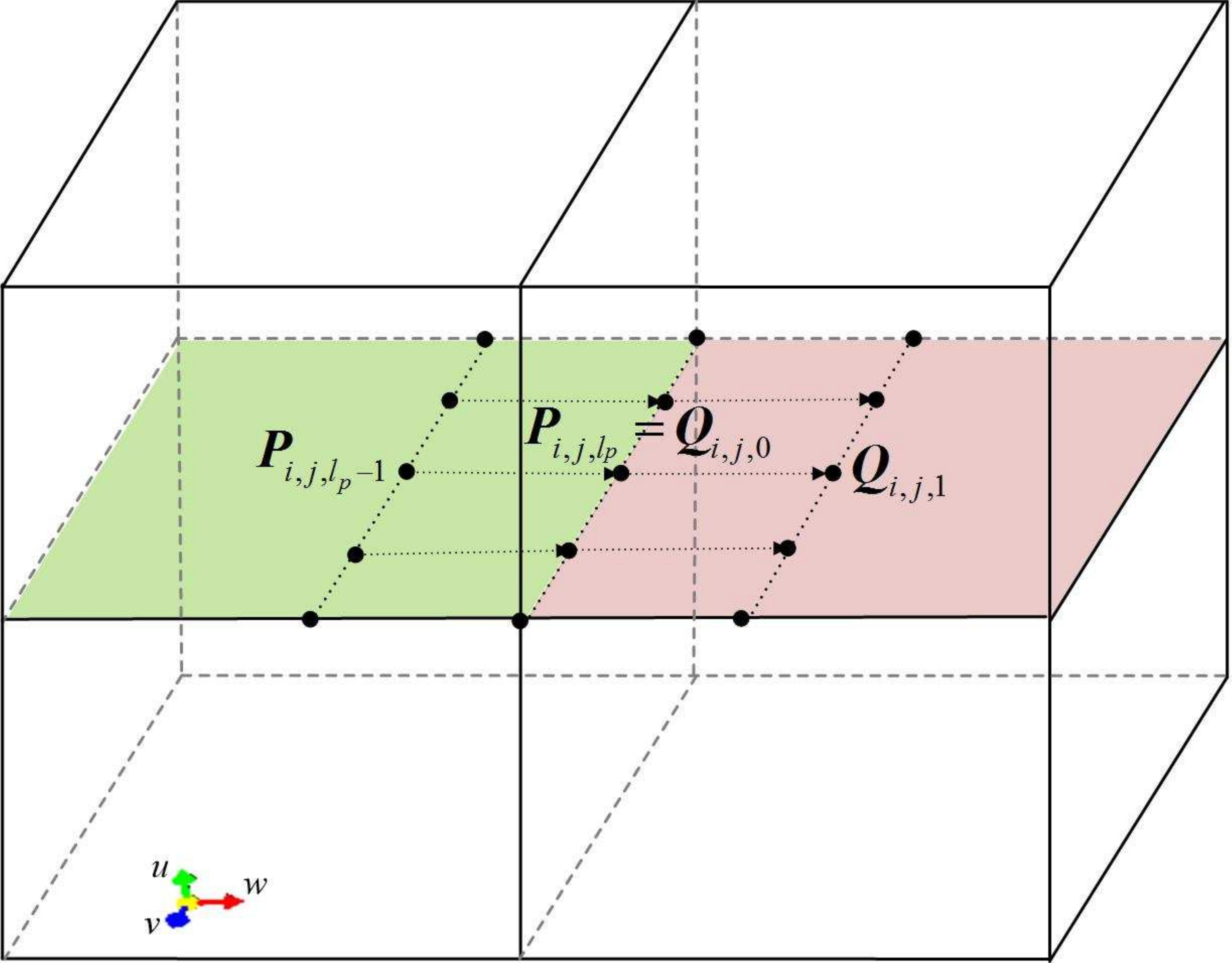}
      \caption{Geometric continuous condition between adjacent TBSs.}
   \label{fig:continuity}
\end{figure}
%------------------------------------------------------------------------------

 %----------------------------------------------------------------------
 % Subsection
 %----------------------------------------------------------------------
 \subsection{Geometric continuity between TBSs}
 \label{subsec:continuity_condition}

 Given two adjacent TBSs (Fig.~\ref{fig:continuity}),
 \begin{equation} \label{eq:tbs_p}
 \begin{split}
   \bm{P}(u,v,w) = & \sum_{i_p=0}^{m_p} \sum_{j_p=0}^{n_p} \sum_{k_p=0}^{l_p}
        \bm{P}_{i_p,j_p,k_p} B_{i_p}(u) B_{j_p}(v) B_{k_p} (w),\\
        & (u,v,w) \in [0,1] \times [0,1] \times [0,1].
 \end{split}
 \end{equation}
 and
 \begin{equation} \label{eq:tbs_q}
 \begin{split}
   \bm{Q}(\mu,\nu,\omega) = & \sum_{i_q=0}^{m_q} \sum_{j_q=0}^{n_q} \sum_{k_q=0}^{l_q}
        \bm{Q}_{i_q,j_q,k_q} B_{i_q}(\mu) B_{j_q}(\nu) B_{k_q} (\omega),\\
        & (\mu,\nu,\omega) \in [0,1] \times [0,1] \times [0,1].
 \end{split}
 \end{equation}
 where, $\bm{P}_{i_p,j_p,k_p}$ and $\bm{Q}_{i_q,j_q,k_q}$ are control points,
 $B_{i_p}(u)$, $B_{j_p}(v)$, $B_{k_p}(w)$, $B_{i_q}(\mu)$, $B_{j_q}(\nu)$, and $B_{k_q}(\omega)$ are B-spline basis,
 the geometric continuity between the two TBSs $\bm{P}(u,v,w)$ and
    $\bm{Q}(\mu,\nu,\omega)$ is defined as,
 %----------------------------------------------------------------------
 % Definition
 %----------------------------------------------------------------------
 \begin{Def}[Geometric Continuity] \label{def:geometric_continuity}
 The two TBSs $\bm{P}(u,v,w)$ and $\bm{Q}(\mu,\nu,\omega)$ are
    $G^n$ geometric continuous along the common boundary surface
    $\bm{P}(u,v,1) = \bm{Q}(\mu,\nu,0)$,
    if,
    \begin{enumerate}
      \item[1)] the two B-spline surfaces $\bm{P}(c,v,w)$ and $\bm{Q}(c,\nu,\omega)$ are $G^n$ geometric continuous along the common boundary curve
          $\bm{P}(c,v,1) = \bm{Q}(c,\nu,0)$, and,
      \item[2)] the two B-spline surfaces $\bm{P}(u,d,w)$ and $\bm{Q}(\mu,d,\omega)$ are $G^n$ geometric continuous along the common boundary curve
          $\bm{P}(u,d,1) = \bm{Q}(\mu,d,0)$,
    \end{enumerate}
    where $c$ and $d$ are arbitrary constants in their domains (refer to Fig.~\ref{fig:continuity}).
 \end{Def}

 The following proposition presents a sufficient condition for the $G^1$
    geometric continuity between two TBSs $\bm{P}(u,v,w)$ and $\bm{Q}(\mu, \nu, \omega)$ along their common boundary surface
    $\bm{P}(u,v,1) = \bm{Q}(\mu,\nu,0)$.
 %----------------------------------------------------------------------
 % Proposition
 %----------------------------------------------------------------------
 \begin{pro} \label{pro:continuity_sufficient}
    Suppose the two TBSs $\bm{P}(u,v,w)$~\pref{eq:tbs_p} and $\bm{Q}(\mu,\nu,\omega)$~\pref{eq:tbs_q} have uniform knot vectors with B\'{e}zier end condition~\cite{farin2002curves}, respectively (Fig.~\ref{fig:continuity}).
    If
    $$\bm{P}_{i,j,l_p} = \bm{Q}_{i,j,0},\ \text{and},\
    \bm{P}_{i,j,l_p}-\bm{P}_{i,j,l_p-1} = \alpha (\bm{Q}_{i,j,1}-\bm{Q}_{i,j,0}),$$
    where $\alpha$ is a positive constant,
    the two TBSs $\bm{P}(u,v,w)$ and $\bm{Q}(\mu,\nu,\omega)$ are $G^1$ geometric continuous along their common boundary surface
    $\bm{P}(u,v,1) = \bm{Q}(\mu,\nu,0)$.
 \end{pro}

 \textbf{Proof:} Referring to Fig.~\ref{fig:continuity},
    if
    $$\bm{P}_{i,j,l_p} = \bm{Q}_{i,j,0},\ \text{and},\
    \bm{P}_{i,j,l_p}-\bm{P}_{i,j,l_p-1} = \alpha (\bm{Q}_{i,j,1}-\bm{Q}_{i,j,0}),$$
    it follows that $\bm{P}'_w(c,v,1) \parallel \bm{Q}'_w(c,\nu,0)$, and
    $\bm{P}'_w(u,d,1) \parallel \bm{Q}'_w(\mu,d,0)$,
    where $c$ and $d$ are arbitrary constants in the domain of $\bm{P}(u,v,w)$ and $\bm{Q}(\mu,\nu,\omega)$.
    That is, the two TBSs $\bm{P}(u,v,w)$ and $\bm{Q}(\mu,\nu,\omega)$ are $G^1$ geometric continuous along  their common boundary surface
    $\bm{P}(u,v,1) = \bm{Q}(\mu,\nu,0)$. $\Box$

 %----------------------------------------------------------------------------
 % Section:
 %----------------------------------------------------------------------------
 \section{Geometric Feasible Direction algorithm}
 \label{sec:gfd}
 As pointed out in Section~\ref{sec:overview}, the composition of TBSs is a step-by-step process, in the order of boundary curve fitting, boundary surface fitting, TBS fitting and smoothness improvement. The fitting problems and smoothness improvement in these steps are all modeled as constraint minimization problems,
    \begin{equation} \label{eq:min_problem}
      \begin{split}
        min_{\bm{P}_i} \quad
             & E(\bm{P}_0, \bm{P}_1, \cdots, \bm{P}_m) \\
        s.t. \quad
             & \text{Constraints}
      \end{split}
    \end{equation}
 where $\bm{P}_i=(x_i,y_i,z_i),\ i=0,1,\cdots,m$ are the unknown control points,
 and $E$ is an objective function.
 Moreover, the conditions for ensuring the validity of curves, surfaces, and TBSs are taken as the hard constraints~\pref{eq:min_problem}. The specific minimization problems will be elucidated in subsequent sections.

 To solve the constraint minimization problem~\pref{eq:min_problem}, we developed the \emph{geometric feasible direction} (GFD algorithm (Algorithm~\ref{alg:gfd}). GFD has clear geometric meanings, so it is a kind of \emph{geometric iterative fitting method}.
 As listed in Algorithm~\ref{alg:gfd}, the inputs to GFD algorithm are an initial B-spline curve, a B-spline surface, or a TBS with control points arranged in a one-dimensional sequence
 $\{\bm{P}^{(0)}_i, i=0,1,\cdots,m\}$,
 and a constrained minimization problem~\pref{eq:min_problem}, including an objective function and the constraints.

 Specifically, in the $k^{th}$ iteration, the gradient vector of the objective function $E$~\pref{eq:min_problem} is,
 \begin{equation} \label{eq:gradient_vector}
 \begin{split}
    \nabla E^{(k)} = \biggl(\frac{\partial E}{\partial x_0}, \frac{\partial E}{\partial y_0}, \frac{\partial E}{\partial z_0}, \cdots, &  \frac{\partial E}{\partial x_i}, \frac{\partial E}{\partial y_i}, \frac{\partial E}{\partial z_i}, \cdots,  \\
    &  \frac{\partial E}{\partial x_m}, \frac{\partial E}{\partial y_m}, \frac{\partial E}{\partial z_m} \biggl)_{\bm{X}^{(k)}},
 \end{split}
 \end{equation}
 where $\bm{X}^{(k)} = (x_0^{(k)},y_0^{(k)},z_0^{(k)}, \cdots, x_i^{(k)},y_i^{(k)},z_i^{(k)}, \cdots, x_m^{(k)},y_m^{(k)},z_m^{(k)})$.
 In the iteration, each control point $\bm{P}^{(k)}_i$ moves along the
    directional vector,
    \begin{equation} \label{eq:direction_vector}
      \bm{D}_i^{(k)} = -\left(\frac{\partial E}{\partial x_i}, \frac{\partial E}{\partial y_i}, \frac{\partial E}{\partial z_i}\right)_{\bm{X}^{(k)}},
    \end{equation}
    to produce the new control point $\bm{P}^{(k+1)}_i$, i.e.,
    \begin{equation*}
      \bm{P}_i^{(k+1)} = \bm{P}_i^{(k)} + \tau_i \bm{D}_i^{(k)},
    \end{equation*}
    where $\tau_i \in [0,1]$ is a weight.
 Initially, we set $\tau_i = 1, i = 0,1,\cdots,m$,
    and the direction $\bm{D}^{(k)}$ in the $k^{th}$ iteration is
    composed of,
    \begin{equation}\label{eq:feasible_direction}
        \bm{D}^{(k)} = (\tau_0 \bm{D}_0^{(k)}, \tau_1 \bm{D}_1^{(k)}, \cdots,
                        \tau_m \bm{D}_m^{(k)}).
    \end{equation}
 Then, the weights $\tau_i, i = 0,1,\cdots,m$ are adjusted one by one to make
    the direction $\bm{D}^{(k)}$~\pref{eq:feasible_direction} a feasible one,
    which satisfies the following two requirements:
    \begin{enumerate}
      \item[1)] The new B-spline curve, surface, or TBS,
    generated by replacing $\bm{P}_i^{(k)}$ with $\bm{P}_i^{(k+1)}$,
    satisfies the constraints of the minimization problem~\pref{eq:min_problem},
      \item[2)]  the condition,
        $-\frac{\bm{D}^{(k)} \cdot \nabla E^{(k)}}
        {\norm{\bm{D}^{(k)}}_2\norm{\nabla E^{(k)}}_2} > \delta_d$,
    holds as well.
    \end{enumerate}
 To this end, we discretize $\tau_i \in [0,1]$ to
    $\{0, \frac{1}{n}, \frac{2}{n}, \cdots, \frac{n}{n}\}$,
    and select a weight as large as possible $\tau_i$, which satisfies the two aforementioned requirements 1) and 2).
 In our implementation, we take $n = 20$, and $\delta_d = 0.3$.

 Moreover, let,
 \begin{equation*}
   \bm{P}^{(k)} = (\bm{P}_0^{(k)}, \bm{P}_1^{(k)}, \cdots, \bm{P}_m^{(k)}).
 \end{equation*}
 After the feasible direction $\bm{D}^{(k)}$ is figured out,
    we calculate the Armijo step $\alpha_k$ (please refer to Ref.~\cite{yaxiang1997theory} for the calculation of the Armijo step).
 Then, the new control points are generated as,
 \begin{equation*}
   \bm{P}^{(k+1)} = \bm{P}^{(k)} + \alpha_k \bm{D}^{(k)}.
 \end{equation*}

 The GFD algorithm is presented in Algorithm~\ref{alg:gfd},
    where the termination condition is taken as,
    \begin{equation}\label{eq:termination}
        \abs{\frac{E(\bm{P}^{(k+1)}) - E(\bm{P}^{(k)})}{E(\bm{P}^{(k)})}} < \varepsilon_e.
    \end{equation}
 The value of $\varepsilon_e$ will be specified in solving specific problems.
 The convergence analysis of the GFD algorithm is demonstrated in the Appendix.

%%------------------------------------------------------------------------------
%% Algorithm: Geometric feasible direction method (GFD)
%%------------------------------------------------------------------------------
%\linesnumbered
\begin{algorithm}[!htb]
\label{alg:gfd}
\caption{Geometric Feasible Direction algorithm}
\KwIn{An initial B-spline curve, B-spline surface, or TBS with control points
    arranged in a one-dimensional sequence
    $\{ \bm{P}_i^{(0)}, i = 0,1,\cdots,m \}$;
    an objective function $E$ and constraints~\pref{eq:min_problem}}
\KwOut{A B-spline curve, B-spline surface, or TBS meeting the termination
    condition}
$k = 0$ \;
\While{the \emph{termination condition} is not reached}{
    %Construct the minimum cone enclosing the difference vectors $\bm{T}_i^{(k)}, i=0,1,\cdots,m-1$~\pref{eq:curve_diff},
%    and calculate its unit axis vector $\bm{T}$   \;
    Calculate the gradient vector $\nabla E^{(k)}$~\pref{eq:gradient_vector} of the objective function $E$~\pref{eq:min_problem} \;
    \For{$i=0$ \KwTo $m$}{
        Determine the directional vector $\bm{D}_i^{(k)}$~\pref{eq:direction_vector} for the control point $\bm{P}_i^{(k)}$ \;
        Calculate the largest possible weight, $\tau_i$, ensuring that requirements 1) and 2) satisfied \;
        }
    Calculate the Armijo step $\alpha_k$~\cite{yaxiang1997theory} \;
    $\bm{P}^{(k+1)} = \bm{P}^{(k)} + \alpha_k \bm{D}^{(k)}$ \;
    $k = k+1$ \;
    %Construct the minimum cone enclosing the difference vectors $\bm{T}_i^{(k)}, i=0,1,\cdots,m-1$ \;
    }
%\end{algorithmic}
\end{algorithm}
 \section{Valid TBS generation by the geometric iterative method}
 \label{sec:algorithm}

 In Section~\ref{sec:overview}, the overview of the method for generating
    valid TBSs was presented.
 In this section, we will elucidate the details of the method.

 %------------------------------------------------------------------------------
%-----------------------------Figure-------------------------------------------
\begin{figure}[!htb]
  \centering
  \subfigure[]{
    \label{subfig:input}
    \includegraphics[width=0.16\textwidth]{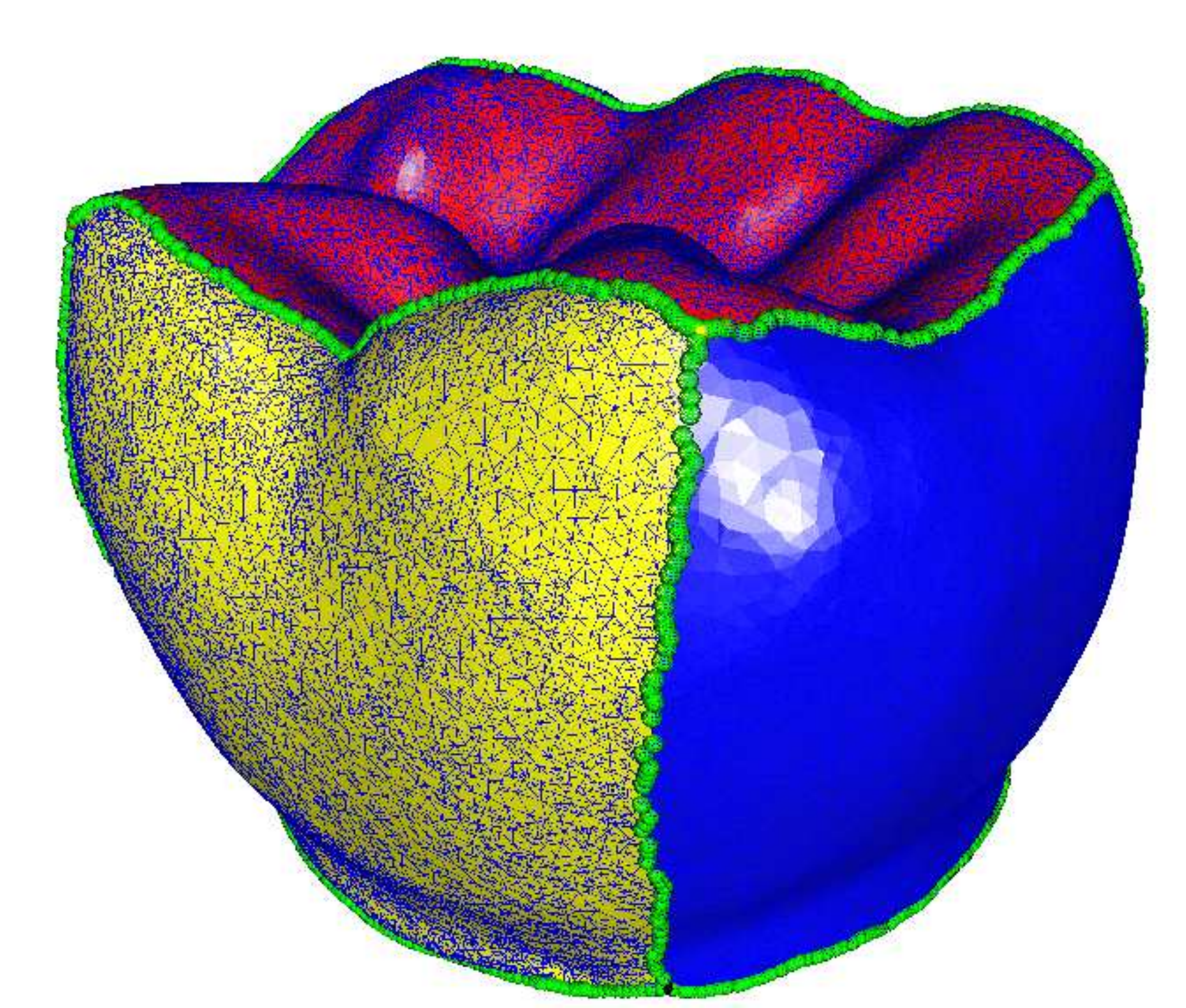}}
  \subfigure[]{
    \label{subfig:para_domain}
    \includegraphics[width=0.13\textwidth]{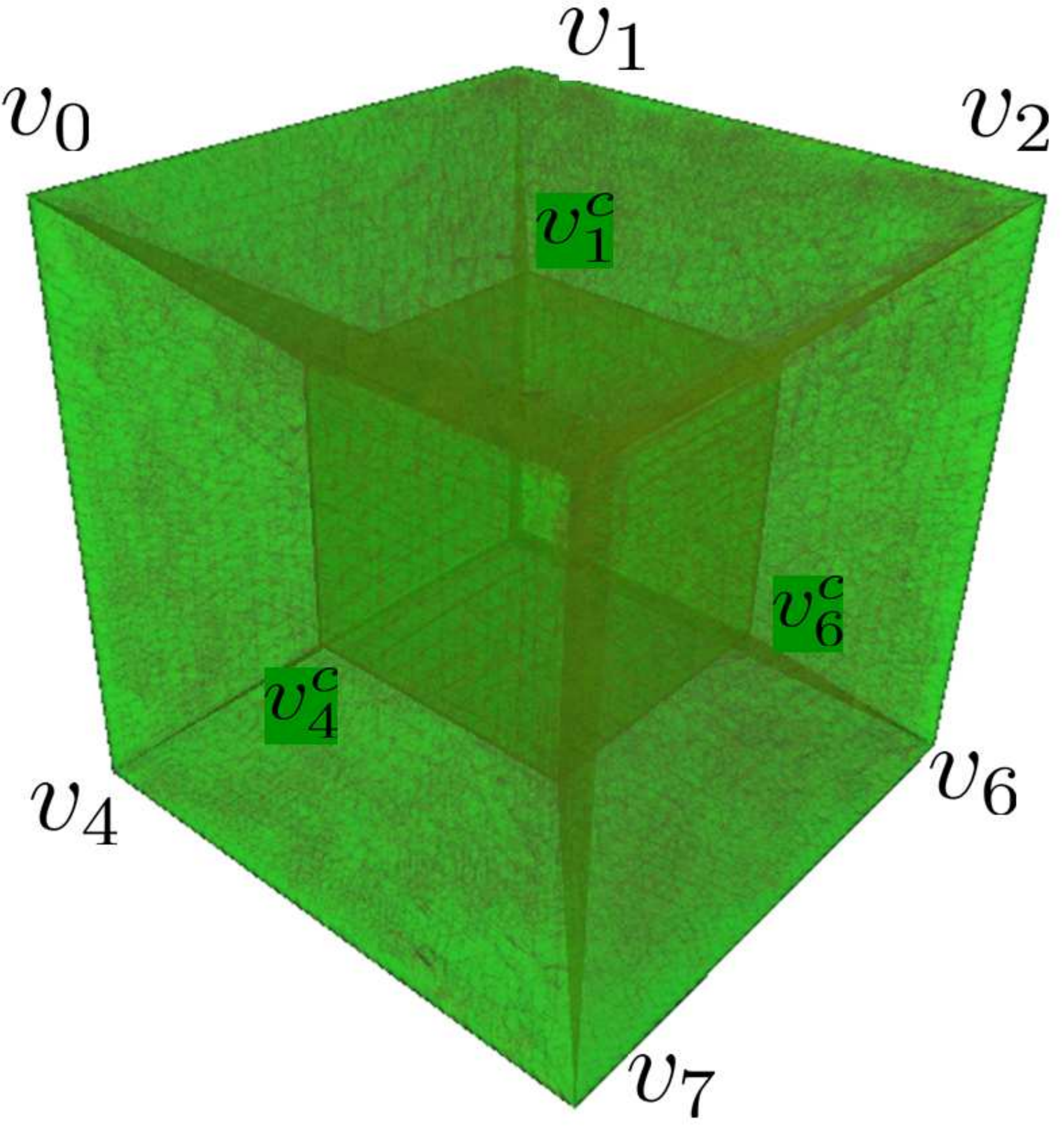}}
  \subfigure[]{
    \label{subfig:divide}
    \includegraphics[width=0.16\textwidth]{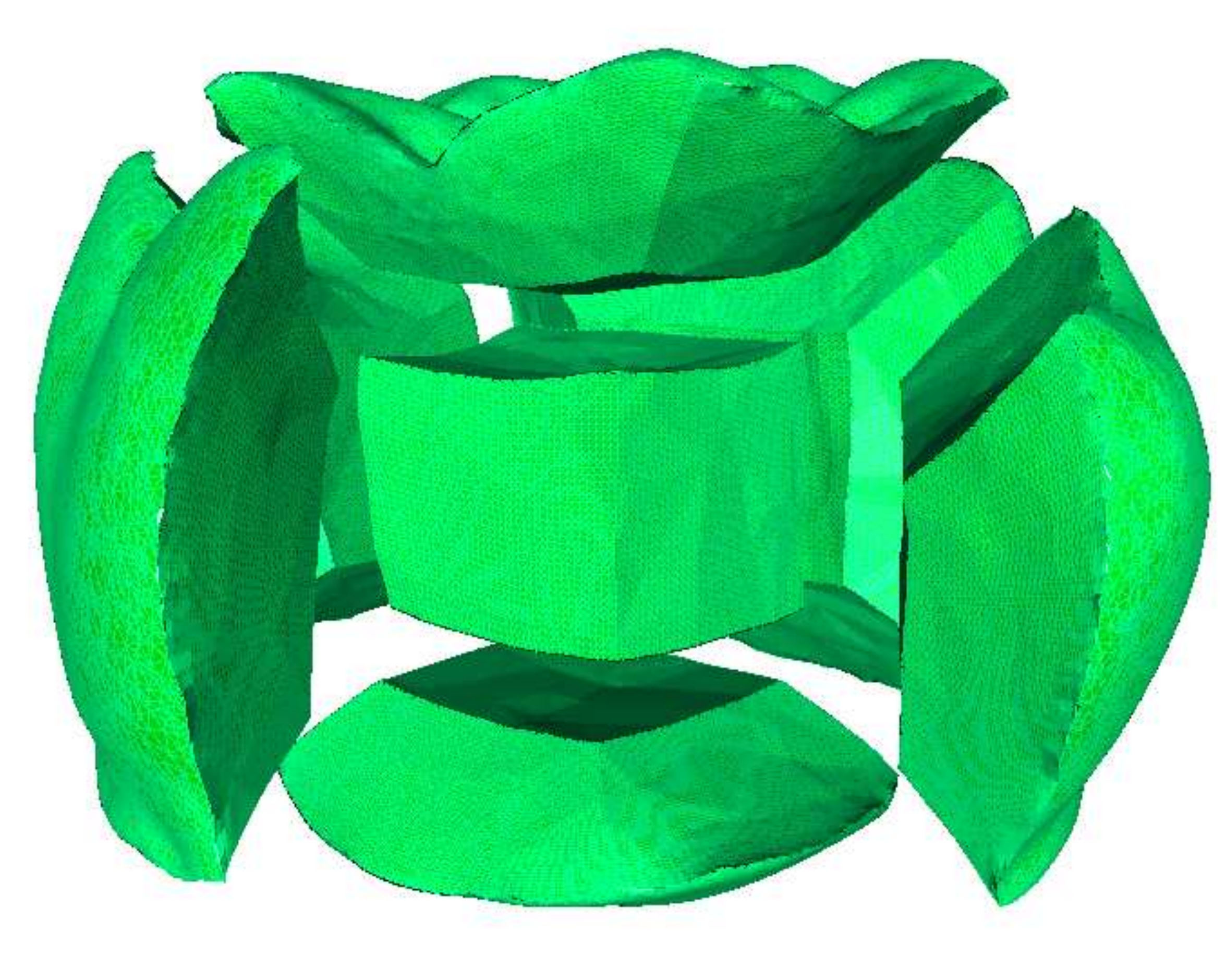}}
      \caption{Partition of the input tet mesh model.
      (a) The input to the developed algorithm is a tet mesh with six surfaces segmented on its boundary mesh.
      (b) The tet mesh is parameterized into the cubic domain $[0,1] \times [0,1] \times [0,1]$, which is partitioned into seven sub-domains.
      (c) Mapping the seven sub-domains into the tet mesh model leads to the seven partitioned sub-volume meshes.}
   \label{fig:input}
\end{figure}
%------------------------------------------------------------------------------

%-----------------------------------------------------------
% Figure
%-----------------------------------------------------------
\begin{figure}[!htb]
  \centering
  \includegraphics[width=0.2\textwidth]{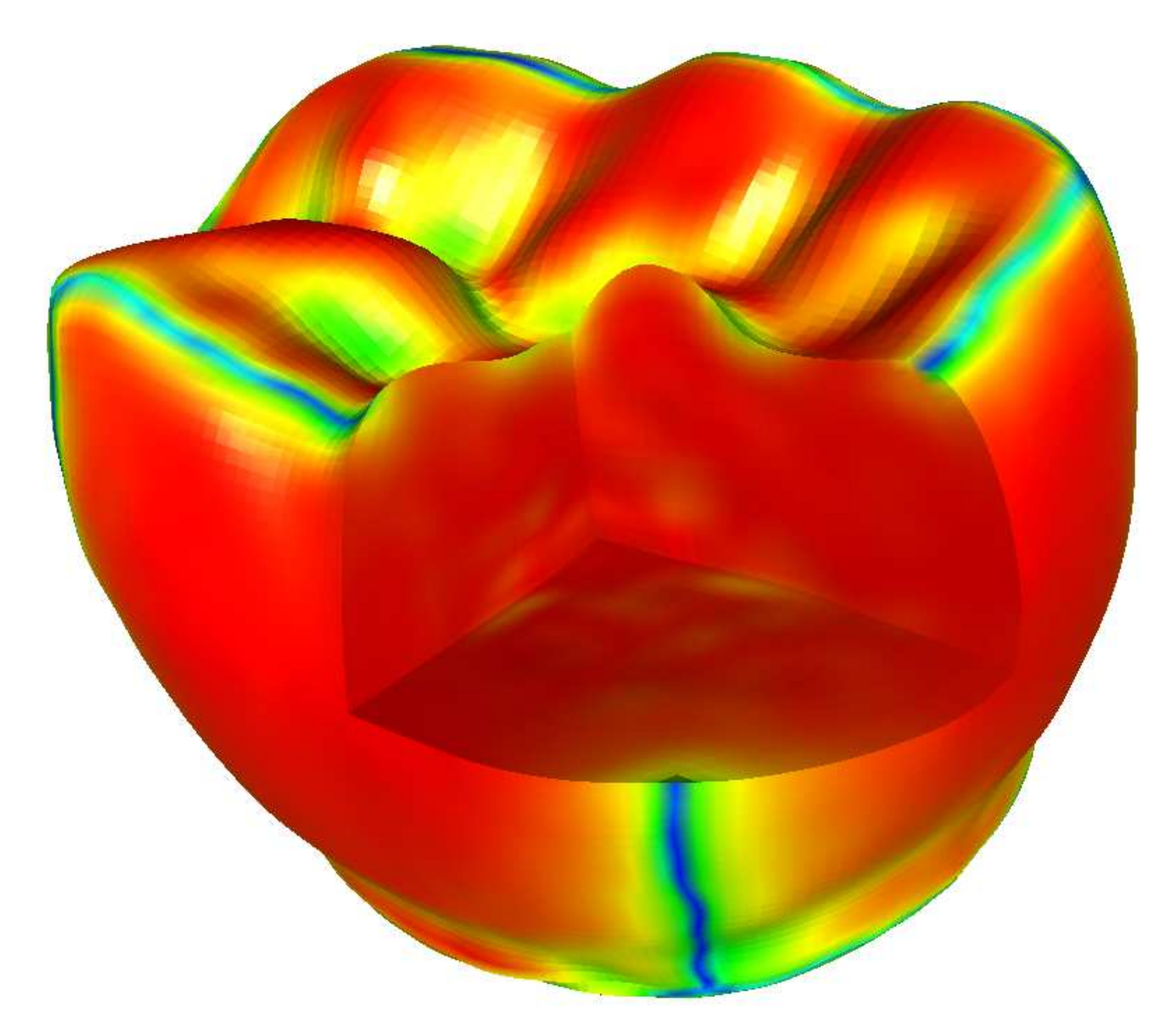}\\
  \caption{The small or negative scaled Jacobian values (in blue) concentrate near the boundaries where the adjacent surfaces are smoothly stitched.}
  \label{fig:negative_jac}
\end{figure}
%-----------------------------------------------------------

 %----------------------------------------------------------------------------
 % Subsection:
 %----------------------------------------------------------------------------
 \subsection{Partition of the tet mesh model by pillow operation}
 \label{subsec:partition}

\begin{figure}[!htb]
  \centering
  \subfigure[]{\label{subfig:quad_before_pillow}
    \includegraphics[width=0.15\textwidth]{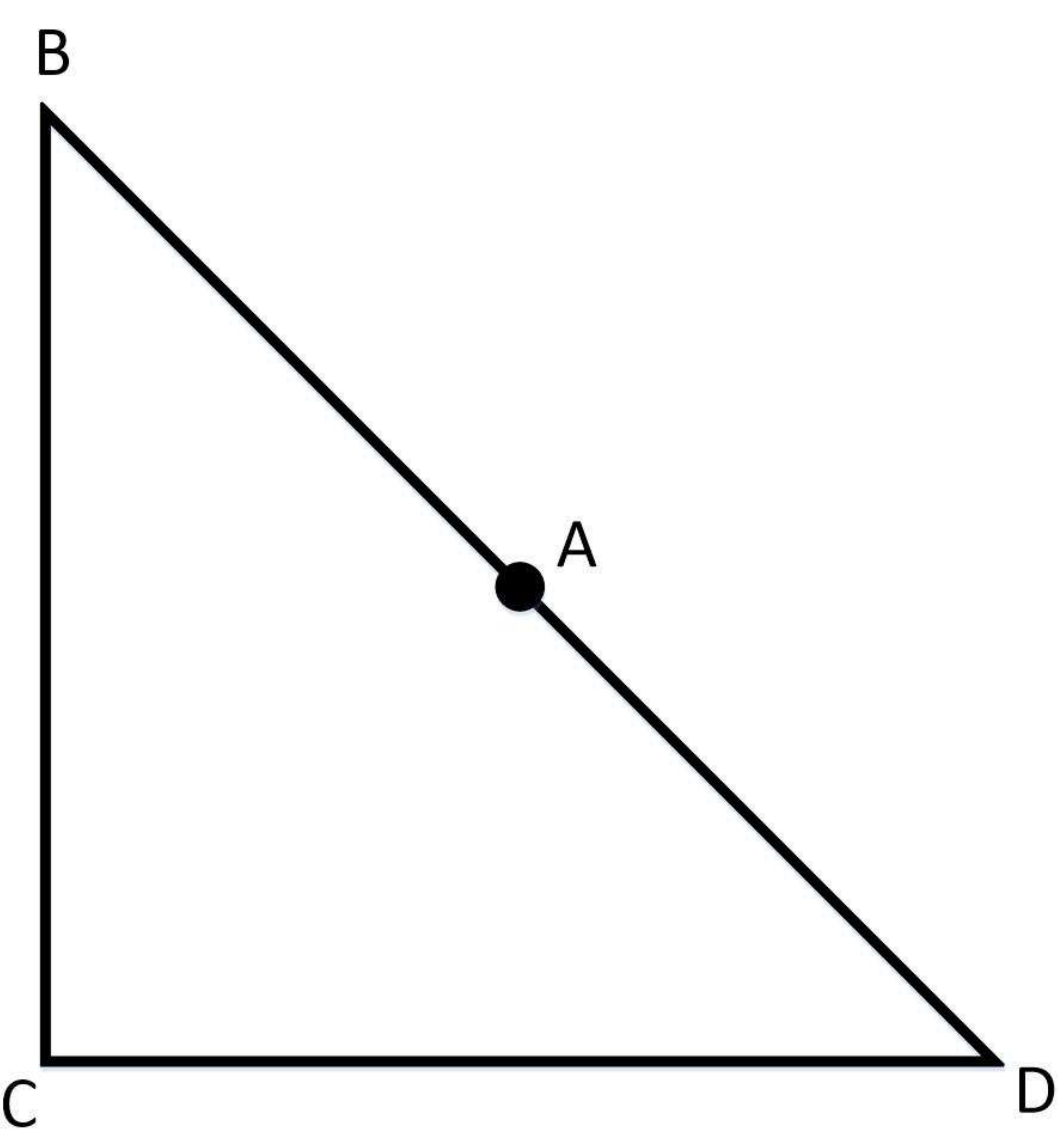}}
  \subfigure[]{\label{subfig:quad_after_pillow}
    \includegraphics[width=0.15\textwidth]{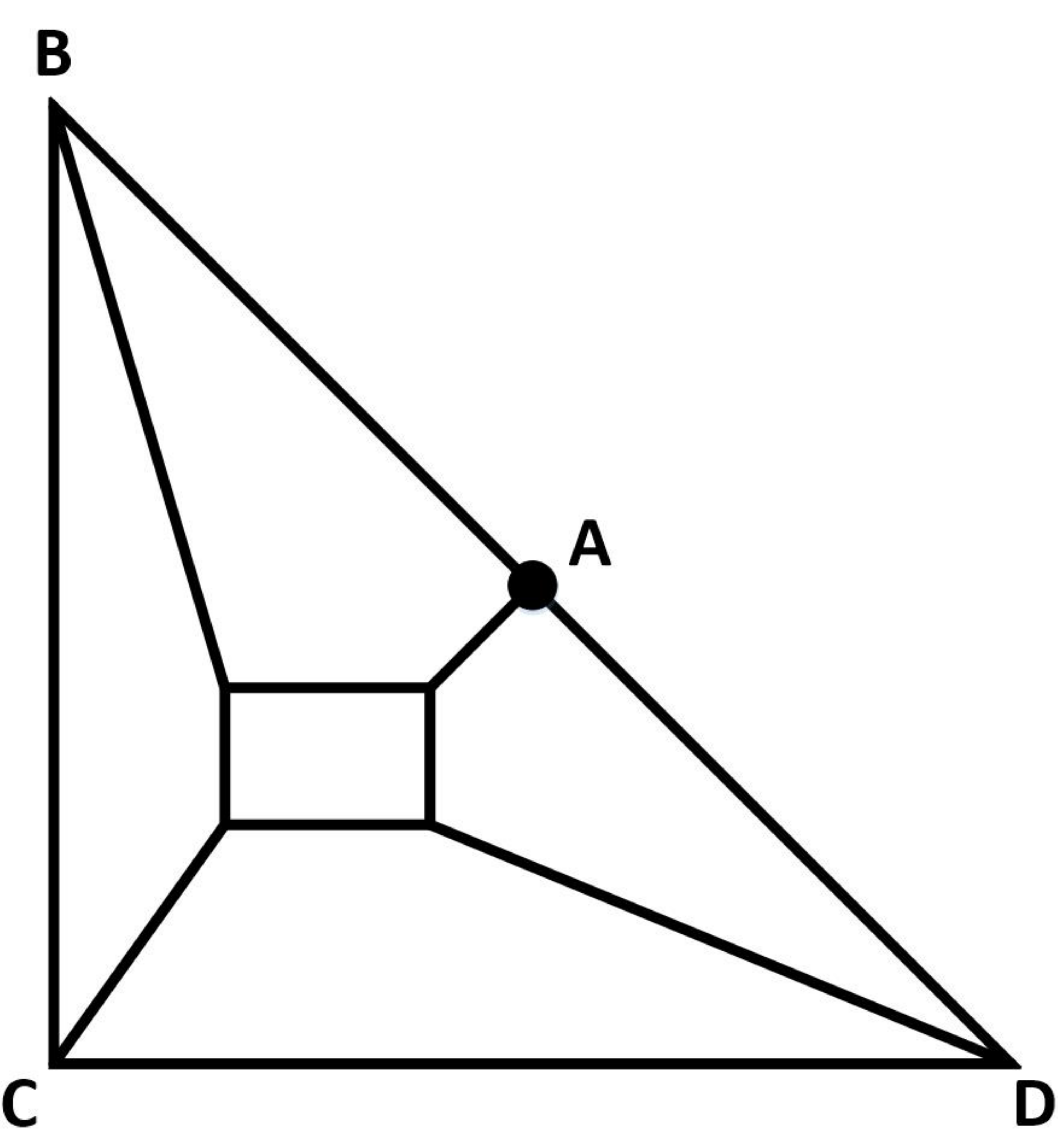}}
      \caption{Improve the mesh quality using the pillow-like operation.
      (a) In the quadrilateral $ABCD$, the Jacobian value at vertex $A$ is $0$, because vertex $D, A$, and $B$ are co-linear.
      (b) By the pillow operation, the Jacobian values at all of the vertices are greater than $0$.}
   \label{fig:pillow}
\end{figure}

 As stated above, the input to our method is a tet mesh model with six
    surfaces segmented on its boundary mesh (Fig.~\ref{subfig:input}).
 The scaled Jacobian value of the generated TBS is heavily influenced by the
    segmentation of the boundary mesh.
 If two adjacent surfaces are $C^n (n \geq 1)$ continuous along their common
    boundary,
    or even concave along the boundary,
    the Jacobian values at the points near the boundaries of the fitting TBS will be small or even negative (Figs.~\ref{fig:negative_jac},~\ref{subfig:quad_before_pillow}).
 This makes the generated TBS invalid.
 Fig.~\ref{fig:negative_jac}illustrates the distribution of the scaled Jacobian
    values of the TBS fitting the tet mesh model in Fig.~\ref{subfig:input} by the method developed in Ref.~\cite{lin2015constructing}.
 It can be seen that, the small or negative Jacobian values mainly
    concentrate in the region (in blue) close to the boundaries of the surfaces, along which two adjacent surfaces are smoothly stitched.

 The quality of the generated TBS can be improved by the pillow
    operation~\cite{mitchell1995pillowing},
    which was originally invented to improve the quality of a hexahedral mesh.
 Refer to Fig.~\ref{subfig:quad_before_pillow},
    there is a quadrilateral $ABCD$,
    and the Jacobian value at vertex $A$ is $0$.
 After performing the pillow operation (Fig.~\ref{subfig:quad_after_pillow}),
    the Jacobian values at all of the vertices are greater than $0$.

%------------------------------------------------------------------------------
%-----------------------------Figure-------------------------------------------
%------------------------------------------------------------------------------

 In order to perform the pillow operation on the input tet mesh,
    the tet mesh model (Fig.~\ref{subfig:input}) is first parameterized into a cubic parameter domain $\Omega = [0,1] \times [0,1] \times [0,1]$ by the volume parameterization method~\cite{lin2015constructing} (Fig.~\ref{subfig:para_domain}).
 Then, a sub-domain
 $$\Omega_c = [\frac{1}{3}, \frac{2}{3}] \times [\frac{1}{3}, \frac{2}{3}] \times [\frac{1}{3}, \frac{2}{3}]$$
    is inserted in the parameter domain $\Omega$.
 As illustrated in Fig.~\ref{subfig:para_domain},
    the vertices of the cubes $\Omega$ and $\Omega_c$
    are denoted as $v_0, v_1, \cdots, v_7$, and $v_0^c, v_1^c, \cdots, v_7^c$, respectively.
 Connecting $v_i$ to $v_i^c, i=0,1,\cdots, 7$ generates six sub-domains
    $\Omega_u, \Omega_d, \Omega_l, \Omega_r, \Omega_f,\ \text{and}\ \Omega_b$.
 For example, the sub-domain $\Omega_u$ is enclosed by the six faces
    $$v_0 v_1 v_2 v_3,\ v_0^c v_1^c v_2^c v_3^c,\ v_0 v_0^c v_1^c v_1,\
    v_1 v_1^c v_2^c v_2,\ v_2 v_2^c v_3^c v_3,\ \text{and},\ v_3 v_3^c v_0^c v_0.$$
 %Meanwhile, the seven sub-domains partition the tet mesh in the parameter
%    domain $\Omega$ into seven tet mesh models.
 Mapping the seven sub-domains into the original tet mesh model produces
    seven partitioned sub-volumes (Fig.~\ref{subfig:divide}).

 %----------------------------------------------------------------------------
 % Subsection:
 %----------------------------------------------------------------------------
 \subsection{Construction of the initial TBSs}
 \label{subsec:initial_tbs}

 In Section~\ref{subsec:partition}, the input tet mesh model is partitioned
    into seven sub-volumes.
 Each of them are parameterized into the cubic parameter domain by the volume
    parameterization method developed in~\cite{lin2015constructing}.
 Note that the parameterization on the common boundary curves and common
    boundary surfaces of adjacent sub-volumes should conform with each other.
 Moreover, each cubic parameter domain is sampled into a
    $(M+1) \times (N+1) \times (K+1)$ grid.
 Similar to the parameterization, the grid on the common boundary curves and
    common boundary surfaces of adjacent sub-volumes should be the same.
 Mapping the grids into the corresponding sub-volumes leads to the control
    grids of the initial TBSs,
    %i.e.,
%    \begin{equation}\label{eq:initial_tbs}
%      \bm{P}^{(0)}(u,v,w) = \sum_{i=0}^M \sum_{j=0}^N \sum_{k=0}^K
%      \bm{P}^{(0)}_{ijk} B_i(u) B_j(v) B_k(w),\
%      (u,v,w) \in [0,1] \times [0,1] \times [0,1],
%    \end{equation}
 whose knot vectors are uniformly distributed in
    $[0,1] \times [0,1] \times [0,1]$ with B\'{e}zier end conditions.
 Owing to the conformity of the parameterization and control grids on the
    common boundary curves and boundary surfaces,
    there is the unique control grid on each boundary curve or boundary surface.

 The geometric iterative fitting method,
    which is employed to fit the sub-volumes (refer to Section~\ref{subsec:iterative_fitting}),
    requires that the initial TBSs should be valid.
 If there is an initial TBS that is not valid,
    the grid sampling rule should be adjusted,
    or the initial TBS should be optimized to produce the valid initial TBS.
 In all of the examples we handled, the initial TBSs generated by the method
    aforementioned are all valid.

 %----------------------------------------------------------------------------
 % Subsection:
 %----------------------------------------------------------------------------
 \subsection{Geometric iterative fitting}
 \label{subsec:iterative_fitting}

 The result generated by the developed method is a composition of seven
    valid TBSs.
 As stated above, the objective function for guaranteeing the validity of
    a TBS is highly nonlinear with a high number of unknowns,
    so the optimization is prone to fail.
 Even if we can find a solution,
    the computation for solving the optimization problem is complicated,
    owing to a significant number of unknowns.
 To reduce the difficulty in guaranteeing the validity of the TBSs,
    we solve this problem step by step,
    in the order of,
    \begin{enumerate}
      \item[(1)] boundary curve fitting (Section~\ref{ssub:curve_fitting}),
      \item[(2)] boundary surface fitting (Section~\ref{ssub:surface_fitting}), and,
      \item[(3)] TBS fitting (Section~\ref{ssub:tbs_fitting}).
    \end{enumerate}
 As mentioned above, the first objective we want to reach is that the
    generated TBSs should be valid,
    that is, the Jacobian value at any point of each TBS should be greater than $0$.
 This means that the boundary curves and boundary surfaces of TBSs should
    also be valid (refer to Section~\ref{subsec:validity_condition} for validity condition).

 %----------------------------------------------------------------------------
 % Subsection:
 %----------------------------------------------------------------------------
 \subsubsection{Boundary curve fitting}
 \label{ssub:curve_fitting}

 Suppose the data point sequence to be fitted is,
 \begin{equation}\label{eq:point_seq}
   \bm{V}_0, \bm{V}_1, \cdots, \bm{V}_{M_c}.
 \end{equation}
 By volume parameterization~\cite{lin2015constructing},
    each data point $\bm{V}_j$ has been assigned a parameter
    $t_j \in [0,1], j=0,1,\cdots,M_c$.
 The data point sequence~\pref{eq:point_seq} will be fitted by a B-spline
    curve,
    \begin{equation} \label{eq:fit_curve}
      \bm{P}(u) = \sum_{i=0}^m \bm{P}_i B_i^k(u),\ u \in [0,1],
    \end{equation}
 where, $\bm{P}_i$ are the control points,
    $B_i^k(u)$ are the B-spline basis functions with degree $d$,
    and the knot vector of the B-spline curve is uniformly defined in $[0,1]$ with B\'{e}zier end conditions.
 %The first order derivative of $\bm{P}(u)$ is,
%    \begin{equation*}
%      \bm{T}(u) = \bm{P}'(u) = (m-k+2)\sum_{i=0}^{m-1} (\bm{P}_{i+1} - \bm{P}_i) B_i^{k-1}(u),\ u \in [0,1].
%    \end{equation*}

 The curve fitting should concern two aspects.
 One is the fitting error; and the other is the validity of
    the fitting curve.
 On one hand, the fitting error can be modeled as,
 \begin{equation*}
   E_{fit}^c = \sum_{j=0}^{M_c} \norm{\bm{P}(u_j) - \bm{V}_j}^2.
 \end{equation*}

 On the other hand, denote $\alpha$ as the aperture of the minimum circular
    cone $\mathcal{C}$ enclosing the difference vectors,
 \begin{equation} \label{eq:curve_diff}
     \left\{ \bm{T}_i = \frac{\bm{P}_{i+1}-\bm{P}_i}{\norm{\bm{P}_{i+1}-\bm{P}_i}},\
     i = 0,1,\cdots,m-1 \right\},
 \end{equation}
 with unit axis vector,
 \begin{equation*}
   \bm{T} = \frac{\sum_{i=0}^{m-1} \bm{T}_i}
                 {\norm{\sum_{i=0}^{m-1} \bm{T}_i}}.
 \end{equation*}
 Evidently, if $\bm{P}'(u) = 0$~\pref{eq:fit_curve},
    the B-spline curve is not valid.
 On the contrary, the larger the norm $\norm{\bm{P}'(u)}$,
    the better is the validity of the B-spline curve~\pref{eq:fit_curve}.
 With fixed lengths of the vectors
    $\bm{T}_i$ in~\pref{eq:curve_diff},
    the smaller the aperture $\alpha$,
    the larger is the norm $\norm{\bm{P}'(u)}$.
 Therefore, the improvement of validity of the fitting curve can be
    achieved by minimizing the following objective function,
    \begin{equation*}
      E_{val}^c = \frac{1}{m} \sum_{i=0}^{m-1}(1-\bm{T}_i \cdot \bm{T}).
    \end{equation*}

 In conclusion, the fitting of the data point sequence~\pref{eq:point_seq}
    with a valid B-spline curve~\pref{eq:fit_curve} can be formulated as the following minimization problem,
    \begin{equation} \label{eq:op_curve}
      \begin{split}
        min_{\bm{P}_i} \quad
             & E^c = (1-\lambda_c) E_{fit}^c + \lambda_c E_{val}^c \\
        s.t. \quad
             & 1)\ \text{The control polygon of}\ \bm{P}(u)\
                   \text{satisfies Proposition~\ref{pro:analysis_curve}},\\
             & 2)\ \text{Compatibility condition},\\
             & 3)\ \bm{P}_0 = \bm{V}_0,\ \bm{P}_m = \bm{V}_{M_c},
      \end{split}
    \end{equation}
    where $\lambda_c \in [0,1]$ is a weight.
 Constraint 1) in the minimization problem~\pref{eq:op_curve} ensures that
    the fitted curve satisfies the validity condition for B-spline curves (Proposition~\ref{pro:analysis_curve}).
 The values of $\lambda_c$ taken in our implementation are listed in
    Table~\ref{tbl:weight}.

%-----------------------------------------------------------
% Figure
%-----------------------------------------------------------
\begin{figure}
  \centering
  \subfigure[]{
  \label{fig:compatibility_curve}
  \includegraphics[width=0.2\textwidth]{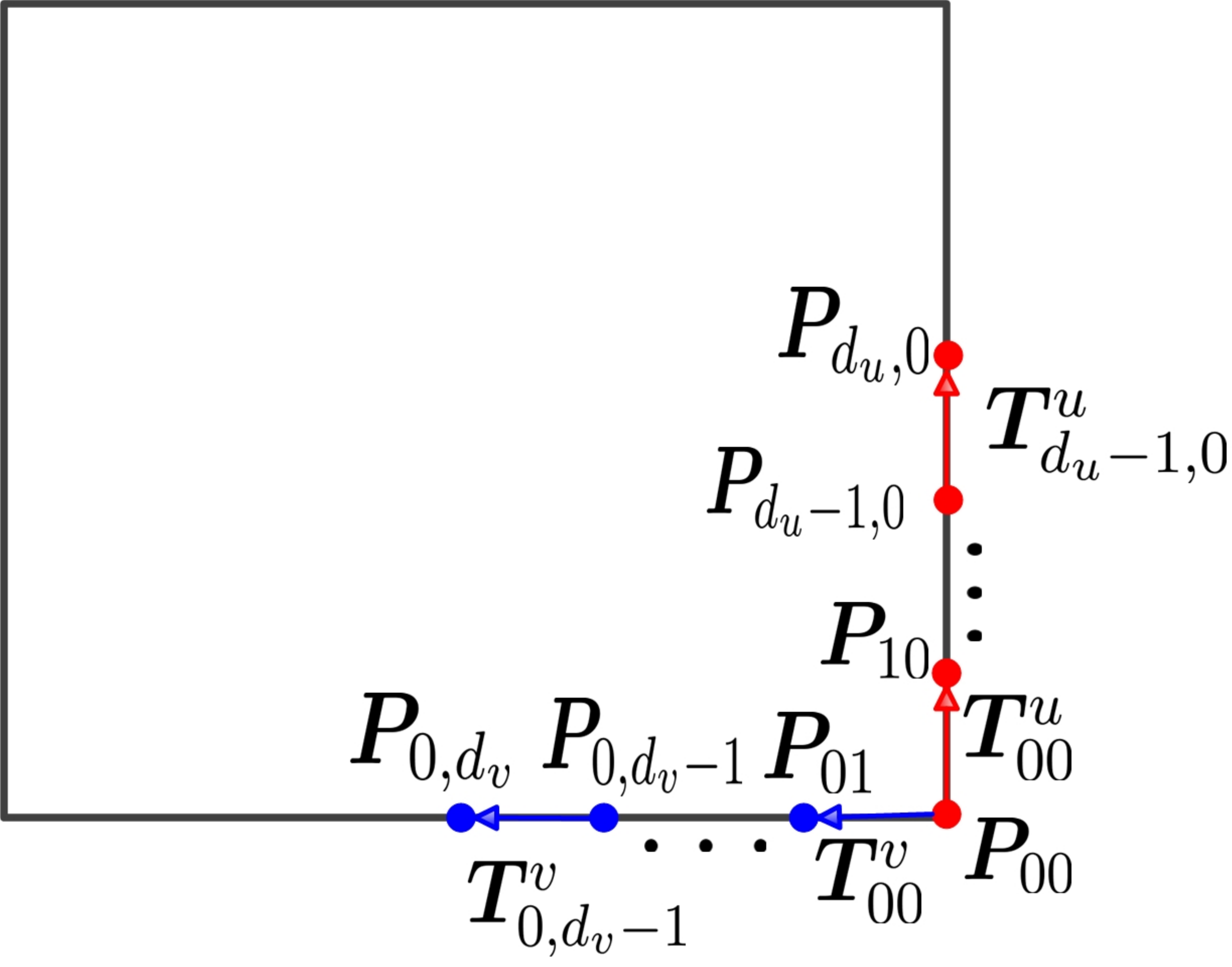}}
  \subfigure[]{
  \label{fig:compatibility_surface}
  \includegraphics[width=0.2\textwidth]{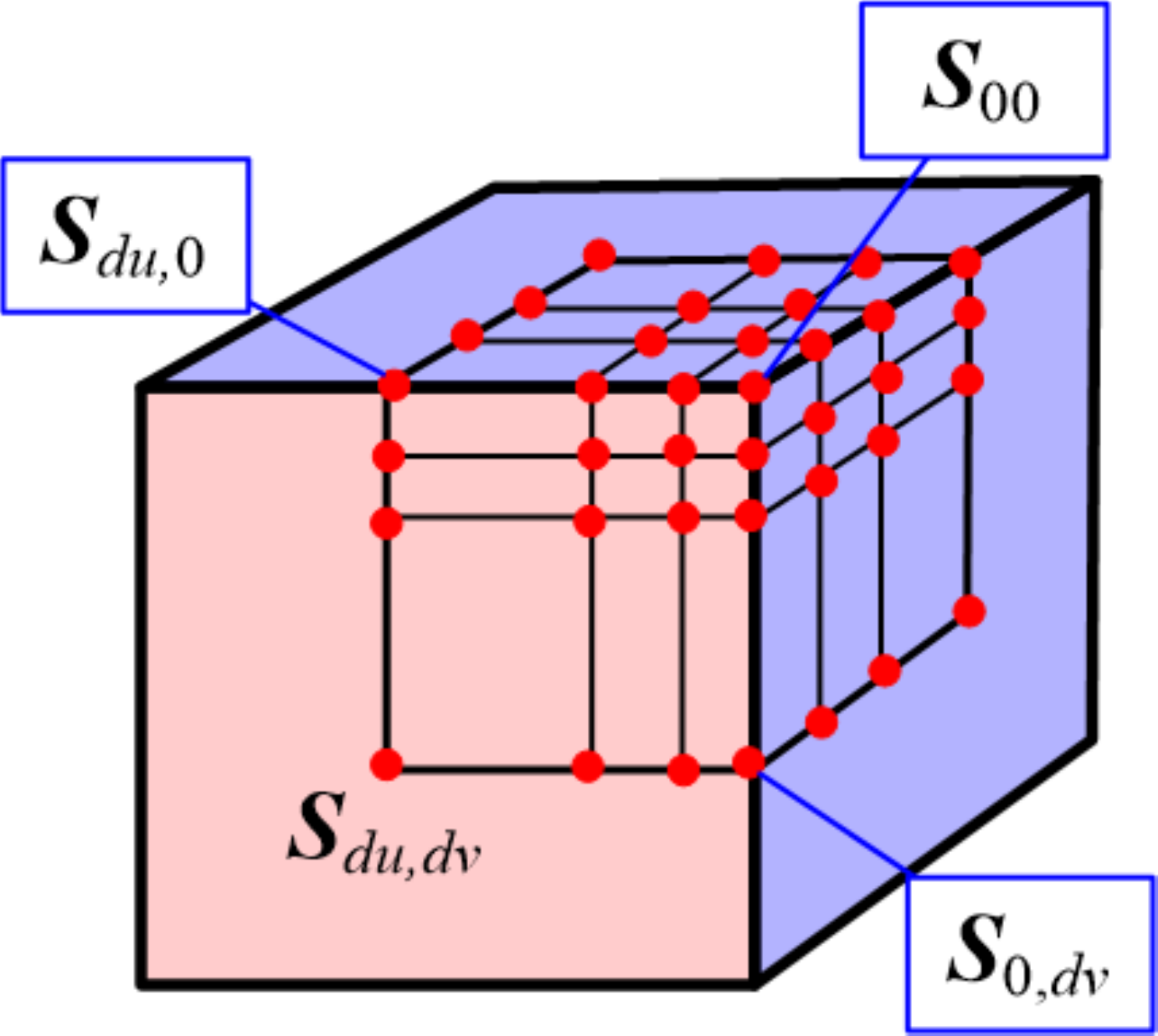}}
  \caption{Compatibility conditions for curve fitting (a) and surface fitting (b).}
  \label{fig:compatibility}
\end{figure}
%-----------------------------------------------------------

 Moreover, the B-spline curves are constructed piece by piece.
 After one piece of B-spline curve is constructed, it is fixed.
 When constructing a new B-spline curve $\bm{P}(u)$,
    it should satisfy the compatibility condition 2) with the adjacent constructed B-spline curves,
    which belong to the same B-spline surface as $\bm{P}(u)$.
 That is, they should satisfy Proposition~\ref{pro:analysis_surface}.
 Specifically, refer to Fig.~\ref{fig:compatibility_curve},
    when constructing the new B-spline curve $\bm{P}(u)$ with control points $\bm{P}_{00}, \bm{P}_{10}, \cdots, \bm{P}_{d_u,0}$ (red points),
    if there exists the piece of B-spline curve adjacent to $\bm{P}_{00}$,
    with control points $\bm{P}_{00}, \bm{P}_{01}, \cdots, \bm{P}_{0,d_v}$ (blue points),
    (the two pieces of B-spline curves belong to the same surface,)
    the difference vectors (refer to Eqs.~\pref{eq:surface_diff_u} and~\pref{eq:surface_diff_v})
    $\bm{T}^u_{00}, \bm{T}^u_{10}, \cdots, \bm{T}^u_{d_u-1,0}$
    and
    $\bm{T}^v_{00}, \bm{T}^v_{01}, \cdots, \bm{T}^v_{0,d_v-1}$
    should satisfy Proposition~\ref{pro:analysis_surface}.

 The constrained minimization problem~\pref{eq:op_curve} is solved by the GFD
    algorithm (Algorithm~\ref{alg:gfd}),
    where the inputs are the initial B-spline curves constructed in Section~\ref{subsec:initial_tbs},
    and the objective function and constraints of the minimization problem~\pref{eq:op_curve}.
 In our implementation, the threshold $\varepsilon_e$ in the termination
    condition~\pref{eq:termination} is taken as $10^{-4}$.

 \subsubsection{Boundary surface fitting}
 \label{ssub:surface_fitting}

 After the boundaries of the seven sub-volumes are fitted by B-spline curves,
    they are fixed.
 The next task is to fit the boundary surfaces of the sub-volumes.
 Each of the boundary surfaces is a triangular mesh with four boundary
    curves.
 Suppose the \emph{inner vertices} of such a triangular mesh are
    $\bm{V}_i, i=0,1,\cdots,M_s$,
    each of which is assigned a pair of parameter values $(u_i,v_i)$ by the volume parameterization~\cite{lin2015constructing}.
 Moreover, suppose the B-spline surface fitted to the triangular mesh is,
 \begin{equation}\label{eq:fit_surface}
   \bm{S}(u,v) = \sum_{i=0}^m \sum_{j=0}^n \bm{S}_{ij} B_i^{d_u}(u) B_j^{d_v}(v),\ (u,v) \in [0,1]\times[0,1],
 \end{equation}
    where, $\bm{S}_{ij}$ are the control points, $B_i^{d_u}(u)$ and $B_j^{d_v}(v)$ are B-spline basis functions with degrees $d_u$ and $d_v$, respectively, both defined in $[0,1]$ with B\'{e}zier end conditions.

 Similar to the curve fitting, the surface fitting also involves two aspects:the fitting precision, and the validity.
 The fitting precision is measured by the following energy function,
 \begin{equation}\label{eq:surface_fit_eng}
   E_{fit}^s = \sum_{k=0}^{M_s} \norm{\bm{S}(u_k,v_k) - \bm{V}_k}^2.
 \end{equation}

 Moreover, the validity of the B-spline surface should be improved as
     per the requirement.
 %Remember that the difference vectors $\bm{T}^u_{ij}$ and $\bm{T}^v_{ij}$
%    along the $u-$ and $v-$direction are defined in Eqs.~\pref{eq:surface_diff_u} and~\pref{eq:surface_diff_v}, respectively.
 Suppose $\mathcal{C}_u$ and $\mathcal{C}_v$ are the minimum circular cones
    enclosing the difference vectors $\bm{T}^u_{ij}$~\pref{eq:surface_diff_u} and $\bm{T}^v_{ij}$~\pref{eq:surface_diff_v}, respectively,
    with unit axis vectors $\bm{T}^u$ and $\bm{T}^v$,
    taken as,
    \begin{equation} \label{eq:surface_axis_vector}
      \bm{T}^u = \frac{\sum_{i=0}^{m-1} \sum_{j=0}^n \bm{T}^u_{ij}}
                {\norm{\sum_{i=0}^{m-1} \sum_{j=0}^n \bm{T}^u_{ij}}},\ \text{and},\
      \bm{T}^v = \frac{\sum_{i=0}^m \sum_{j=0}^{n-1} \bm{T}^v_{ij}}
                {\norm{\sum_{i=0}^m \sum_{j=0}^{n-1} \bm{T}^v_{ij}}}.
    \end{equation}
 The validity of a B-spline surface can be improved by two strategies.
 Firstly, as analyzed in Section~\ref{ssub:curve_fitting},
    the apertures of the two cones should be as small as possible,
    which is achieved by minimizing the following two energy functions,
 \begin{align*}
   E^s_u & = \frac{1}{m(n+1)}\sum_{i=0}^{m-1} \sum_{j=0}^n \left( 1- \bm{T}_{ij}^u \cdot \bm{T}^u \right),\ \text{and},\\
   E^s_v & = \frac{1}{(m+1)n}\sum_{i=0}^m \sum_{j=0}^{n-1} \left( 1- \bm{T}_{ij}^v \cdot \bm{T}^v \right).
 \end{align*}
 Secondly, the two cones should be as close to perpendicularity as possible,
    which is formulated as the minimization of the following energy function (Eq.~\pref{eq:surface_axis_vector}),
    \begin{equation*}
      E^s_{uv} = (\bm{T}^u \cdot \bm{T}^v)^2.
    \end{equation*}

 Therefore, the B-spline surface fitting the triangular mesh with validity
    guarantee can be generated by the following minimization problem,
 \begin{equation} \label{eq:op_surface}
      \begin{split}
        & min_{\bm{S}_{ij}}  \quad E^s = (1-\lambda_s-\mu_s) E_{fit}^s + \lambda_s (E_u^s + E_v^s) + \mu_s E_{uv}^s \\
        s.t.\ &  1)\ \text{The control polygon of}\
               \bm{S}(u,v)\ \text{satisfies Proposition~\ref{pro:analysis_surface}}, \\
             &  2)\ \text{Compatibility condition},\ \text{and}, \\
             &  3)\ \text{The boundary control points of}\
               \bm{S}(u,v)~\pref{eq:fit_surface}\ \text{are fixed},
      \end{split}
    \end{equation}
 %\begin{equation} \label{eq:op_surface}
%      \begin{split}
%        & min_{\bm{S}_{ij}} \quad E^s = (1-\lambda_s-\mu_s) E_{fit}^s + \lambda_s (E_u^s + E_v^s) + \mu_s E_{uv}^s \\
%        & s.t.\ 1)\ \bm{T}_{ij}^u \cdot \bm{T}^u > \bm{T}_{ij}^u \cdot \bm{T}^v, i=0,1,\cdots,m-1, j = 0,1,\cdots, n,\\
%        & \quad 2)\ \bm{T}_{ij}^v \cdot \bm{T}^v > \bm{T}_{ij}^v \cdot \bm{T}^u, i=0,1,\cdots,m, j = 0,1,\cdots, n-1,\\
%        & \quad 3)\ \text{The control points of the boundary curves are fixed as generated in Section~\ref{ssub:curve_fitting}},
%      \end{split}
%    \end{equation}
    where $\lambda_s, \mu_s \in [0,1]$ are weights,
    whose values can be found in Table~\ref{tbl:weight},
    and Constraint 1) in Eq.~\pref{eq:op_surface} ensures that the B-spline
    surface $\bm{S}(u,v)$~\pref{eq:fit_surface} satisfies the validity condition for B-spline surfaces (Proposition~\ref{pro:analysis_surface}).

%%-----------------------------------------------------------
%% Figure
%%-----------------------------------------------------------
%\begin{figure}
%  % Requires \usepackage{graphicx}
%  \centering
%  \includegraphics[width=0.2\textwidth]{../pic/compatibility_patch.eps}\\
%  \caption{Compatibility condition in Eq.~\pref{eq:op_surface}.}
%  \label{fig:compatibility_surface}
%\end{figure}
%%-----------------------------------------------------------

 Similar to the curve case, the B-spline surfaces are constructed
     piece by piece. Hence,
    the construction of a new B-spline surface $\bm{S}(u,v)$
    (the surface in light red in Fig.~\ref{fig:compatibility_surface}),
    should satisfy the \emph{compatibility condition} 2) (Eq.~\pref{eq:op_surface}) with existing B-spline surfaces (the two surfaces in light blue in Fig.~\ref{fig:compatibility_surface}),
    which are adjacent to one of the corners of the new B-spline surface $\bm{S}(u,v)$,
    and belong to the same TBS with $\bm{S}(u,v)$.
 In other words,
    the corresponding sub-control-polygons of these surfaces adjacent to one corner of $\bm{S}(u,v)$ should satisfy Proposition~\ref{pro:analysis_tbs}.
 Specifically, refer to Fig.~\ref{fig:compatibility_surface},
    supposing the degree of the TBS is $d_u \times d_v \times d_w$,
    %if the degrees of these surfaces are $d_u \times d_v$, $d_v \times d_w$,
%    and $d_u \times d_w$, respectively,
    the three sub-control-polygons with size $d_u \times d_v$, $d_v \times d_w$, and $d_u \times d_w$, respectively (refer to Eq.~\pref{eq:sub_control_polygon}),
    should satisfy the validity condition for TBSs, i.e., Proposition~\ref{pro:analysis_tbs}.

 The constrained minimization problem~\pref{eq:op_surface} is solved by the
    GFD algorithm (Algorithm~\ref{alg:gfd}).
 The inputs to the GFD algorithm are the initial B-spline surfaces
    constructed in Section~\ref{subsec:initial_tbs},
    the objective function and constraints of the minimization problem~\pref{eq:op_surface}.
 According to Eq.~\pref{eq:fit_surface},
    the control points of the B-spline surface are arranged in a one-dimensional sequence with lexicographic order, i.e.,
    $
      \{\bm{S}_{00}, \bm{S}_{01}, \cdots, \bm{S}_{mn} \}.
    $
 It should be pointed out that the threshold $\varepsilon_e$ in the
    termination condition~\pref{eq:termination} is set as $10^{-5}$.

 %Similar as the curve fitting case, the constrained minimization
%    problem~\pref{eq:op_surface} is solved by the GFD algorithm (refer to Algorithm~\ref{alg:ggd}).
% The initial B-spline surfaces are the surfaces generated in
%    Section~\ref{subsec:initial_tbs} with boundary B-spline curves re-fitted in Section~\ref{ssub:curve_fitting}.
%
% Details of the GFD algorithm can be found in Algorithm~\ref{alg:ggd}.
% Specifically, in step 3, the objective function is $E^s$~\pref{eq:op_surface}.
% In step 5, the directional vector $\bm{D}_{ij}^{(k)}$ for the control point
%    $\bm{S}_{ij}^{(k)} = (x_{ij}^{(k)}, y_{ij}^{(k)}, z_{ij}^{(k)})$ is,
%    \begin{equation*}
%      \bm{D}_{ij}^{(k)} = \left(\frac{\partial E^s}{\partial x_{ij}}, \frac{\partial E^s}{\partial y_{ij}}, \frac{\partial E^s}{\partial z_{ij}} \right)_{X^{(k)}},
%    \end{equation*}
%    where $X^{(k)} = (x_{00}^{(k)},y_{00}^{(k)},z_{00}^{(k)}, \cdots, x_{mn}^{(k)},y_{mn}^{(k)},z_{mn}^{(k)})$,
%    arranged in lexicographic order.
% Finally, in step 6, we will calculate an as large as possible weight $\tau$ making $\bm{S}_{ij}^{(k)} - \tau \bm{D}_{ij}^{(k)}$ satisfy the constraints 1) and 2) in Eq.~\pref{eq:op_surface}.

 %----------------------------------------------------------------------------
 % Subsection:
 %----------------------------------------------------------------------------
 \subsubsection{TBS fitting}
 \label{ssub:tbs_fitting}

 Now, it is the time to fit seven sub-volumes by TBSs.
 Similar to the boundary curve fitting and boundary surface fitting,
    the seven sub-volumes are also fitted one by one.
 Each sub-volume is a tetrahedral (tet) mesh,
    whose boundary triangular meshes were fitted with B-spline surfaces (refer to Section~\ref{ssub:surface_fitting}),
    and thus fixed in the TBS fitting.
 Suppose $\bm{V}_{n}, n=0,1,\cdots,M_h$ are the inner vertices of a
    sub-volume mesh with parameter values $(u_n,v_n,w_n)$,
    and the TBS fitting the sub-volume mesh vertices is,
    \begin{equation} \label{eq:fit_tbs}
      \bm{H}(u,v,w) = \sum_{i=0}^{m_u} \sum_{j=0}^{m_v} \sum_{k=0}^{m_w}
      \bm{H}_{ijk} B_i^{d_u}(u) B_j^{d_v}(v) B_k^{d_w}(w),
    \end{equation}
    where $\bm{H}_{ijk}$ are the control points,
    $B_i^{d_u}(u), B_j^{d_v}(v)$, and $B_k^{d_w}(w)$ are the B-spline basis functions of degrees $d_u, d_v$, and $d_w$,
    defined on the interval $[0,1]$,
    with B\'{e}zier end conditions, respectively.

 Similar to the boundary curve and surface fitting,
    the generation of a TBS should involve the following factors:
     \begin{enumerate}
       \item[1)] The fitting precision to the tet mesh vertices, and,
       \item[2)] the improvement of validity of each TBS.
       %\item[3)] the smoothness between two adjacent TBSs, and inside each TBS.
     \end{enumerate}
 On one hand, the fitting precision is modeled by the following formula,
 \begin{equation}\label{eq:tbs_fit_eng}
   E^h_{fit} = \sum_{n=0}^{M_h} \norm{\bm{H}(u_n,v_n,w_n) - \bm{V}_n}^2.
 \end{equation}
 On the other hand, to improve the validity of a TBS,
    we define the difference vectors along the $u-, v-$, and $w-$directions, respectively, as,
    \begin{equation*}
    \begin{split}
      \bm{T}^u_{ijk} & = \frac{\bm{H}_{i+1,j,k} - \bm{H}_{ijk}}
                            {\norm{\bm{H}_{i+1,j,k} - \bm{H}_{ijk}}},\quad
      \bm{T}^v_{ijk} = \frac{\bm{H}_{i,j+1,k} - \bm{H}_{ijk}}
                            {\norm{\bm{H}_{i,j+1,k} - \bm{H}_{ijk}}},\\
      \bm{T}^w_{ijk} & = \frac{\bm{H}_{i,j,k+1} - \bm{H}_{ijk}}
                            {\norm{\bm{H}_{i,j,k+1} - \bm{H}_{ijk}}}.
    \end{split}
    \end{equation*}
 %According to Proposition~\ref{pro:analysis_tbs}, if,
% \begin{equation*}
%   \bm{T}^u_{l;i_u,j_u,k_u} \cdot (\bm{T}^v_{l;i_v,j_v,k_v} \times \bm{T}^w_{l;i_w,j_w,k_w}) > 0,
% \end{equation*}
%    the TBS is valid.
 Suppose $\mathcal{C}_u, \mathcal{C}_v$, and $\mathcal{C}_{w}$ are the
    minimum circular cones enclosing
    $\bm{T}^u_{ijk}$, $\bm{T}^v_{ijk}$, and $\bm{T}^w_{ijk}$,
    with unit axis vectors $\bm{T}^u, \bm{T}^v$, and $\bm{T}^w$, taken as,
    respectively,
    \begin{equation} \label{eq:tbs_axis_vector}
      \bm{T}^u = \frac{\sum_{ijk}\bm{T}^u_{ijk}}
                      {\norm{\sum_{ijk}\bm{T}^u_{ijk}}},\
      \bm{T}^v = \frac{\sum_{ijk}\bm{T}^v_{ijk}}
                      {\norm{\sum_{ijk}\bm{T}^v_{ijk}}},\
      \bm{T}^w = \frac{\sum_{ijk}\bm{T}^w_{ijk}}
                      {\norm{\sum_{ijk}\bm{T}^w_{ijk}}}.
    \end{equation}
 Similar as the analysis in Section~\ref{ssub:surface_fitting},
    the smaller the apertures of the cones,
    the better the validity of the TBS,
    which is modeled as the minimization of the following three energy functions, i.e.,
    \begin{equation*}
    \begin{split}
      E^h_u & = \frac{1}{N_u} \sum_{i,j,k} (1 - \bm{T}^u_{ijk} \cdot \bm{T}^u),\quad
      E^h_v = \frac{1}{N_v} \sum_{i,j,k} (1 - \bm{T}^v_{ijk} \cdot \bm{T}^v),\\
      E^h_w & = \frac{1}{N_w} \sum_{i,j,k} (1 - \bm{T}^w_{ijk} \cdot \bm{T}^w),
    \end{split}
    \end{equation*}
    where, $N_u, N_v$, and $N_w$ are the number of the vectors $\bm{T}^u_{ijk}, \bm{T}^v_{ijk}$, and $\bm{T}^w_{ijk}$, respectively.
    In addition, the closer the perpendicularities of the three cones are to each other, the better is the validity of the TBS, which can be formulated as the minimization of the following energy functions:
    \begin{equation*}
      E^h_{uv} = (\bm{T}^u \cdot \bm{T}^v)^2,\quad
      E^h_{vw} = (\bm{T}^v \cdot \bm{T}^w)^2,\quad
      E^h_{uw} = (\bm{T}^u \cdot \bm{T}^w)^2.
    \end{equation*}

 Essentially, the minimization problem for generating the valid
    TBSs can be modeled as,
 \begin{equation} \label{eq:op_tbs}
      \begin{split}
        & min_{\bm{H}_{ijk}} \quad E^h = (1-\lambda_h-\mu_h) E_{fit}^h + \lambda_h (E_u^h + E_v^h + E_w^h) \\
        & \qquad \qquad  + \mu_h (E_{uv}^h + E_{uw}^h + E_{vw}^h) \\
        s.t. & \quad 1)\ \text{The control grid of $\bm{H}(u,v,w)$
                satisfies Proposition~\ref{pro:analysis_tbs}},\\
             &  \quad 2)\ \text{The boundary control points of
                $\bm{H}(u,v,w)$~\pref{eq:fit_tbs} are fixed},
      \end{split}
    \end{equation}
    where $\lambda_h, \mu_h \in [0,1]$ are weights,
    and Constraint 1) makes the TBS $\bm{H}(u,v,w)$ satisfy the validity condition for TBSs (Proposition~\ref{pro:analysis_tbs}).
 The values of weights $\lambda_h, \mu_h \in [0,1]$ are presented in
        Table~\ref{tbl:weight}.

 The constrained minimization problem~\pref{eq:op_tbs} is solved
    by the GFD algorithm (Algorithm~\ref{alg:gfd}),
    with the initial TBSs constructed in Section~\ref{subsec:initial_tbs} as input.
 The control points of an input TBS (refer to Eq.~\pref{eq:fit_tbs}) are
    arranged in a one-dimensional sequence with lexicographic order, i.e.,
    $\{\bm{H}_{000}, \bm{H}_{001}, \cdots, \bm{H}_{m_u,m_v,m_w}\}$.
 Moreover, we set $\varepsilon_e = 10^{-6}$ in the termination
    condition~\pref{eq:termination}.

 %The minimization problem~\pref{eq:op_tbs} is solved using the GFD
%    algorithm with minor differences from Algorithm~\ref{alg:ggd}.
% The input to the GFD algorithm for TBS fitting are the initial TBSs
%    generated in Section~\ref{subsec:initial_tbs},
%    with boundary surfaces re-fitted in Section~\ref{ssub:surface_fitting}.
% Moreover, in step 3, the objective function is $E^h$~\pref{eq:op_tbs}.
% In Step 5, the directional vector $\bm{D}_{ijk}^{(l)}$ for the control point
%    $\bm{H}_{ijk}^{(l)}$ is,
%    \begin{equation*}
%      \bm{D}_{ijk}^{l} = \left( \frac{\partial E^h}{\partial x},
%                                \frac{\partial E^h}{\partial y}, \frac{\partial E^h}{\partial z}
%                                \right)_{X^{(l)}},
%    \end{equation*}
%    where $X^{(l)} = (x_{000}^{(l)}, y_{000}^{(l)}, z_{000}^{(l)}, \cdots,
%                      x_{m_u, m_v, m_w}^{(l)}, y_{m_u,m_v,m_w}^{(l)},z_{m_u,m_v,m_w}^{(l)})$,
%    arranged in lexicographic order.
% In step 6, an as large as possible weight $\tau$ is calculated to make
%    $\bm{H}_{ijk}^{(l)} - \tau \bm{H}_{ijk}^{(l)}$ satisfy the constraints 1) in Eq.~\pref{eq:op_tbs}.

 %----------------------------------------------------------------------------
 % Subsection:
 %----------------------------------------------------------------------------
 \subsubsection{Smoothness and fairness improvement}
 \label{ssub:tbs_smooth}

 In this section, we will improve the smoothness between two adjacent
    TBSs,
    and the fairness of each TBS.
 Suppose $\bm{H}_l(u,v,w), l = 0,1,\cdots,6$ is a TBS with control points
    $\bm{H}_{l,ijk}$.
 Since the boundary paths are fixed before the TBS fitting,
    the two adjacent TBSs have reached $G^0$ continuity.
 %The $G^1$ continuity between the two adjacent TBSs is guaranteed by
%    Proposition~\ref{pro:continuity_sufficient}.
 According to Proposition~\ref{pro:continuity_sufficient} and
    Fig.~\ref{fig:continuity},
    the two adjacent TBSs $H_{l_1}(u,v,w)$ and $H_{l_2}(u,v,w)$ are $G^1$ continuous,
    if the two unit vectors $\bm{T}_{l_1;ij}$ and $\bm{T}_{l_2;ij}$, i.e.,
 %For example, in the case illustrated in Fig.~\ref{fig:continuity},
 \begin{equation*}
   \bm{T}_{l_1;ij} = \frac{\bm{P}_{i,j,l_p}-\bm{P}_{i,j,l_p-1}}
                          {\norm{\bm{P}_{i,j,l_p}-\bm{P}_{i,j,l_p-1}}},\ \text{and},\
   \bm{T}_{l_2;ij} = \frac{\bm{Q}_{i,j,1}-\bm{Q}_{i,j,0}}
                          {\norm{\bm{Q}_{i,j,1}-\bm{Q}_{i,j,0}}},
 \end{equation*}
    have the same direction.
 Then, the smoothness between a pair of adjacent TBSs can be improved by
    the minimization of the energy function,
    \begin{equation*}
      \sum_{ij} (1 - \bm{T}_{l_1;ij} \cdot \bm{T}_{l_2;ij}).
    \end{equation*}
 The smoothness between all of the adjacent TBSs can be
    improved by minimizing the energy,
    \begin{equation*}
      E^h_{smooth} = \sum_{\substack{\text{any pair of adjacent TBSs}\\ H_{l_1}\ \text{and}\ H_{l_2}}} \sum_{i,j} (1 - \bm{T}_{l_1; i j} \cdot \bm{T}_{l_2;i j}).
    \end{equation*}
 Meanwhile, the fairness of the TBSs is improved by minimizing the fairness
    energy,
    \begin{equation*}
    \begin{split}
      E^h_{fair} = & \sum_l \int_0^1 \int_0^1 \int_0^1
      \left( \left( \frac{\partial^2 \bm{H}_l}{\partial u^2} \right)^2 +
             \left( \frac{\partial^2 \bm{H}_l}{\partial v^2} \right)^2 +
             \left( \frac{\partial^2 \bm{H}_l}{\partial w^2} \right)^2 +
             \right.\\
      & \left.
       2\left(\frac{\partial^2 \bm{H}_l}{\partial u \partial v}\right)^2 +
       2\left(\frac{\partial^2 \bm{H}_l}{\partial u \partial w}\right)^2 +
       2\left(\frac{\partial^2 \bm{H}_l}{\partial v \partial w}\right)^2      \right) du dv dw. \\
    \end{split}
    \end{equation*}
 The fairness energy $E^h_{fair}$ is a quadratic function of control points
    of the TBS $H_l$.
 In our implementation, the composite trapezoidal
    rule~\cite{burden2001numerical} is utilized to calculate the integrals in the fairness energy $E^h_{fair}$.

 Therefore, the improvement of smoothness and fairness with
    validity guarantee can be modeled as the constrained minimization problem,
 \begin{equation} \label{eq:op_smooth}
      \begin{split}
        & min_{\bm{H}_{l;ijk}} \quad (1-\lambda_f)E^h_{smooth} +
                                    \lambda_f E^h_{fair} \\
        s.t.\ & 1)\ \text{The control grid of each TBS
                satisfies Proposition~\ref{pro:analysis_tbs}},\\
            & 2)\ \text{The boundary control points of each TBS are fixed},
      \end{split}
    \end{equation}
 where, $\lambda_f \in [0,1]$ is a weight,
    listed in Table~\ref{tbl:weight}.
 Constraint 1) in~\pref{eq:op_smooth} guarantees
    that each TBS satisfies the validity condition for TBSs (Proposition~\ref{pro:analysis_tbs}).
 %Moreover, based on Proposition~\ref{pro:continuity_sufficient},
%    the variables $\bm{H}_{l;ijk}$ in the minimization problem~\pref{eq:op_smooth} are just the control points adjacent to the boundary control points of each TBS.

 The constrained minimization problem~\pref{eq:op_smooth} is solved by the
    GFD algorithm (Algorithm~\ref{alg:gfd}).
 The input to the GFD algorithm is the composition of the seven TBSs,
    constructed in Section~\ref{ssub:tbs_fitting},
    with control points arranged in a one-dimensional sequence.
 The threshold $\varepsilon_e$ in the termination
    condition~\pref{eq:termination} is taken as $10^{-5}$.

\begin{figure}[!htb]
  \centering
  \subfigure[]{ \label{subfig:mannequin_hex}
    \includegraphics[width=0.20\textwidth]{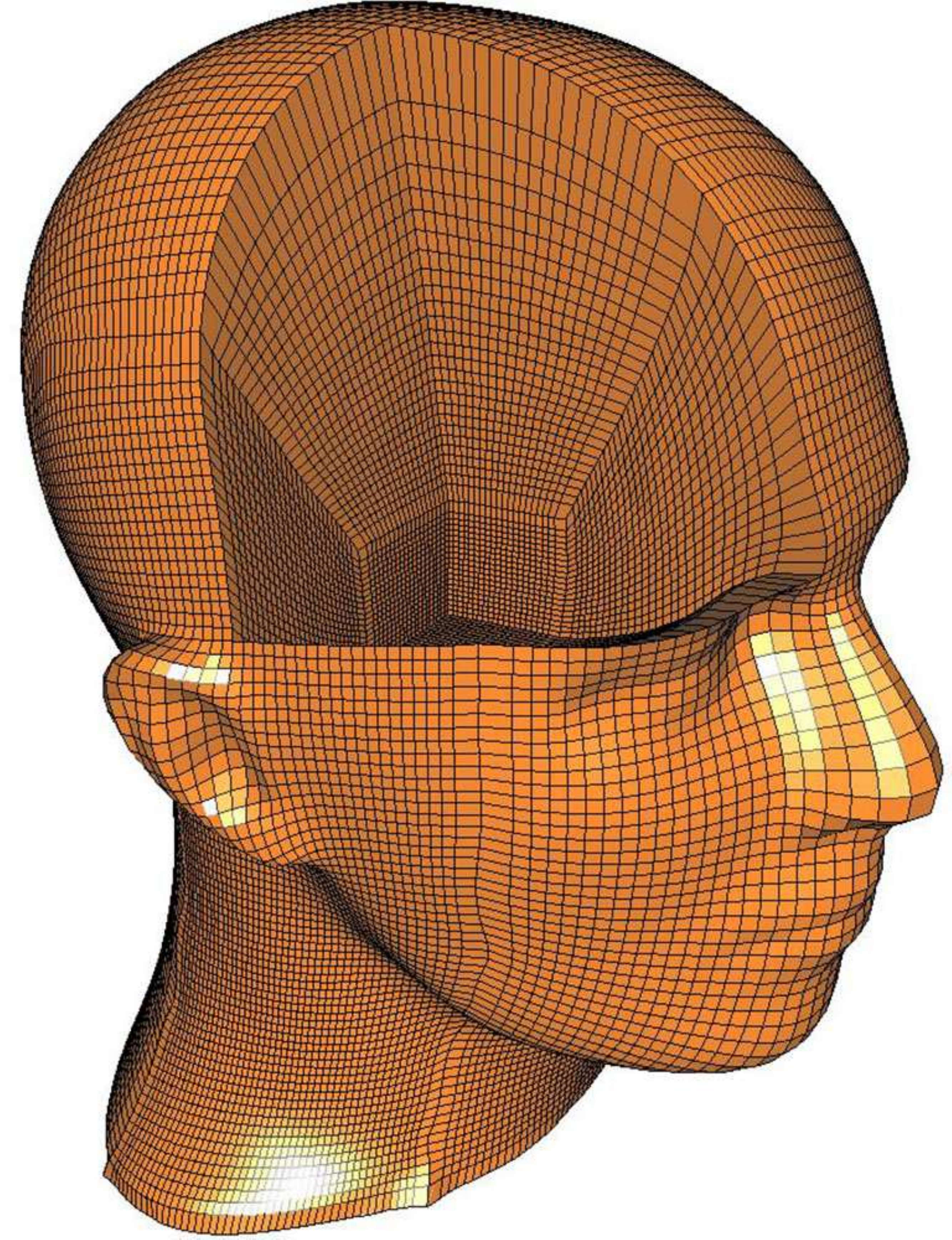}}
  \subfigure[]{ \label{subfig:teeth_hex}
    \includegraphics[width=0.25\textwidth]{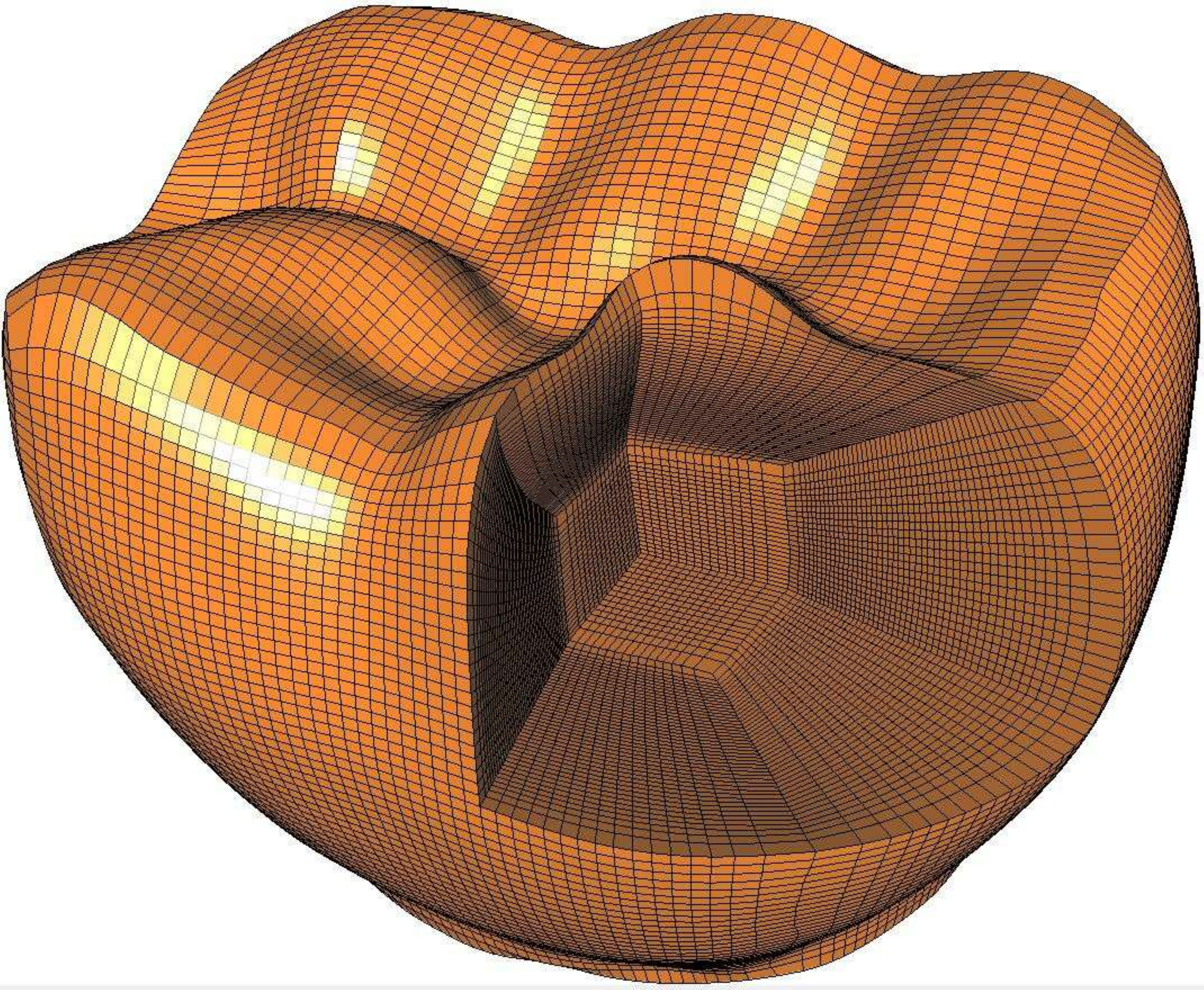}}
  \subfigure[]{ \label{subfig:duck_hex}
    \includegraphics[width=0.24\textwidth]{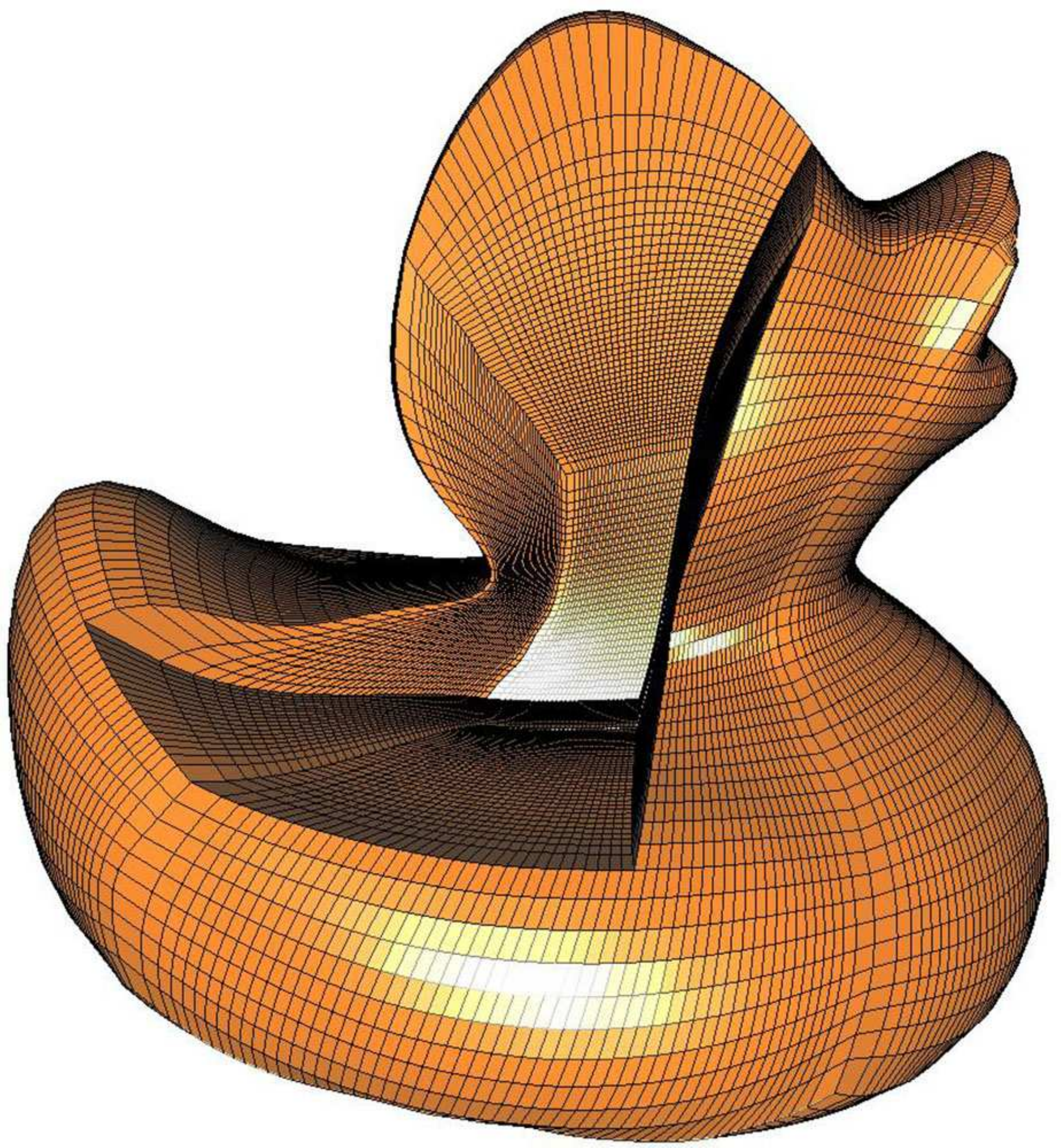}}
  \subfigure[]{ \label{subfig:moai_hex}
    \includegraphics[width=0.17\textwidth]{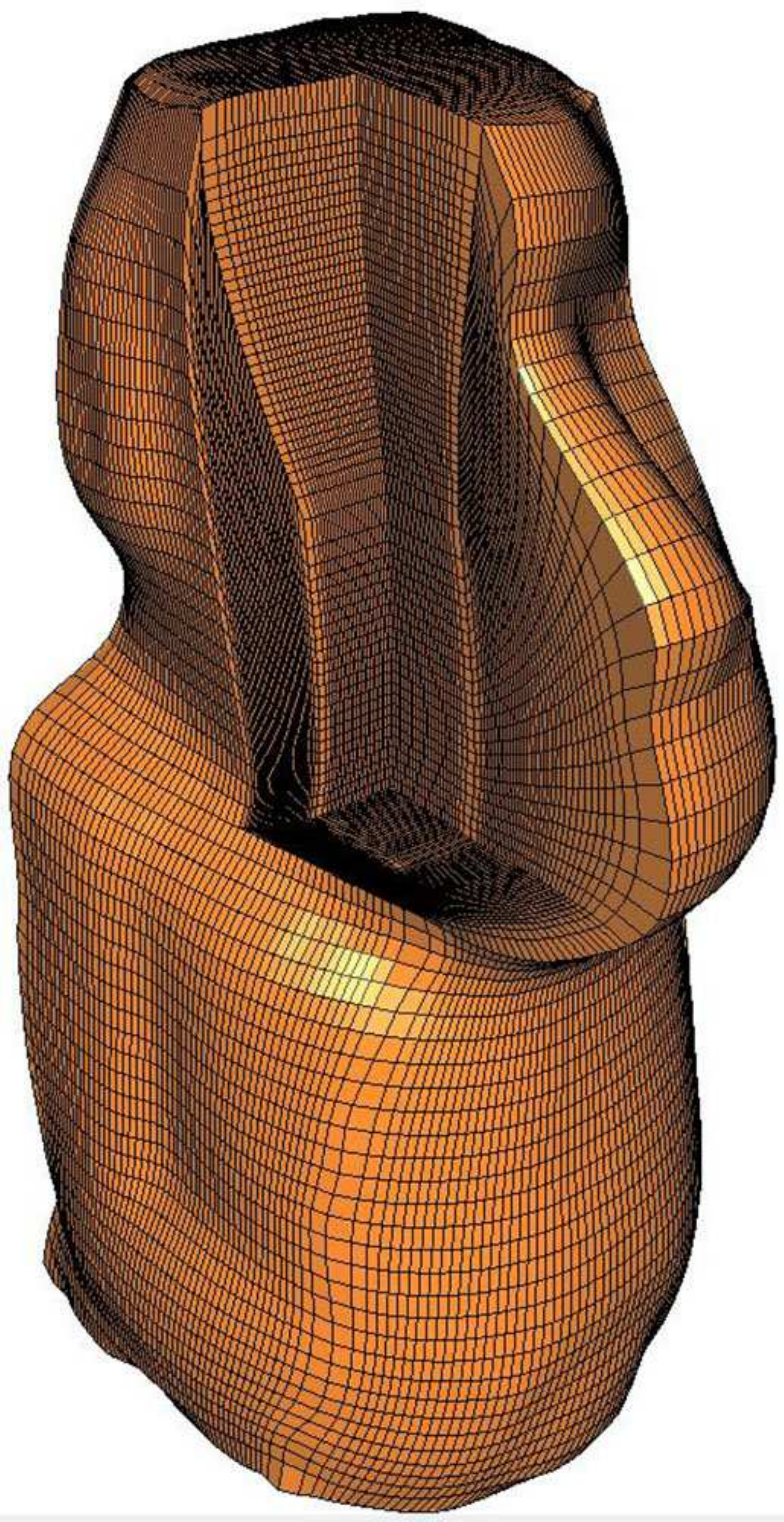}}
  \subfigure[]{ \label{subfig:ball_joint_hex}
    \includegraphics[width=0.23\textwidth]{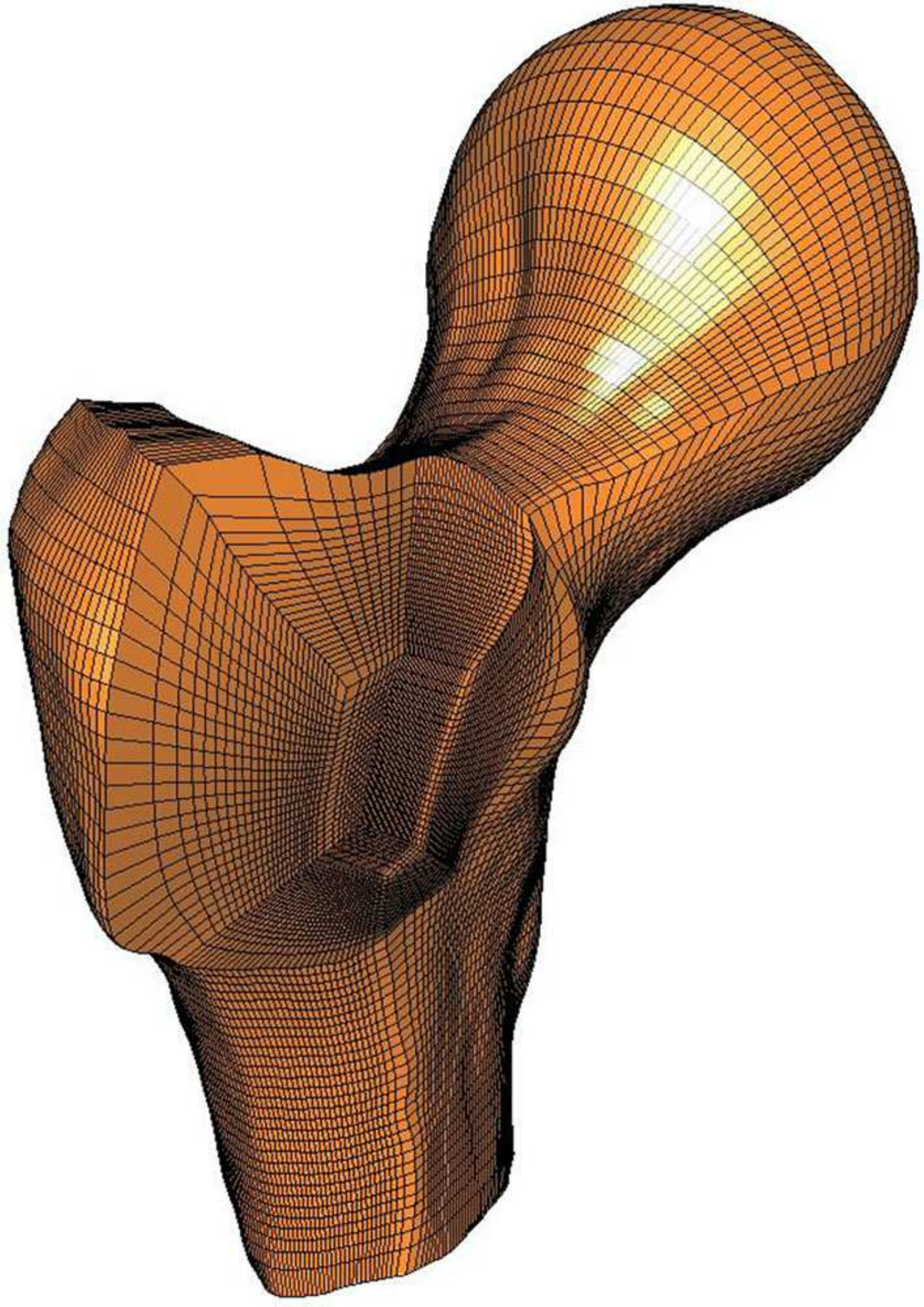}}
  \subfigure[]{ \label{subfig:venus_hex}
    \includegraphics[width=0.21\textwidth]{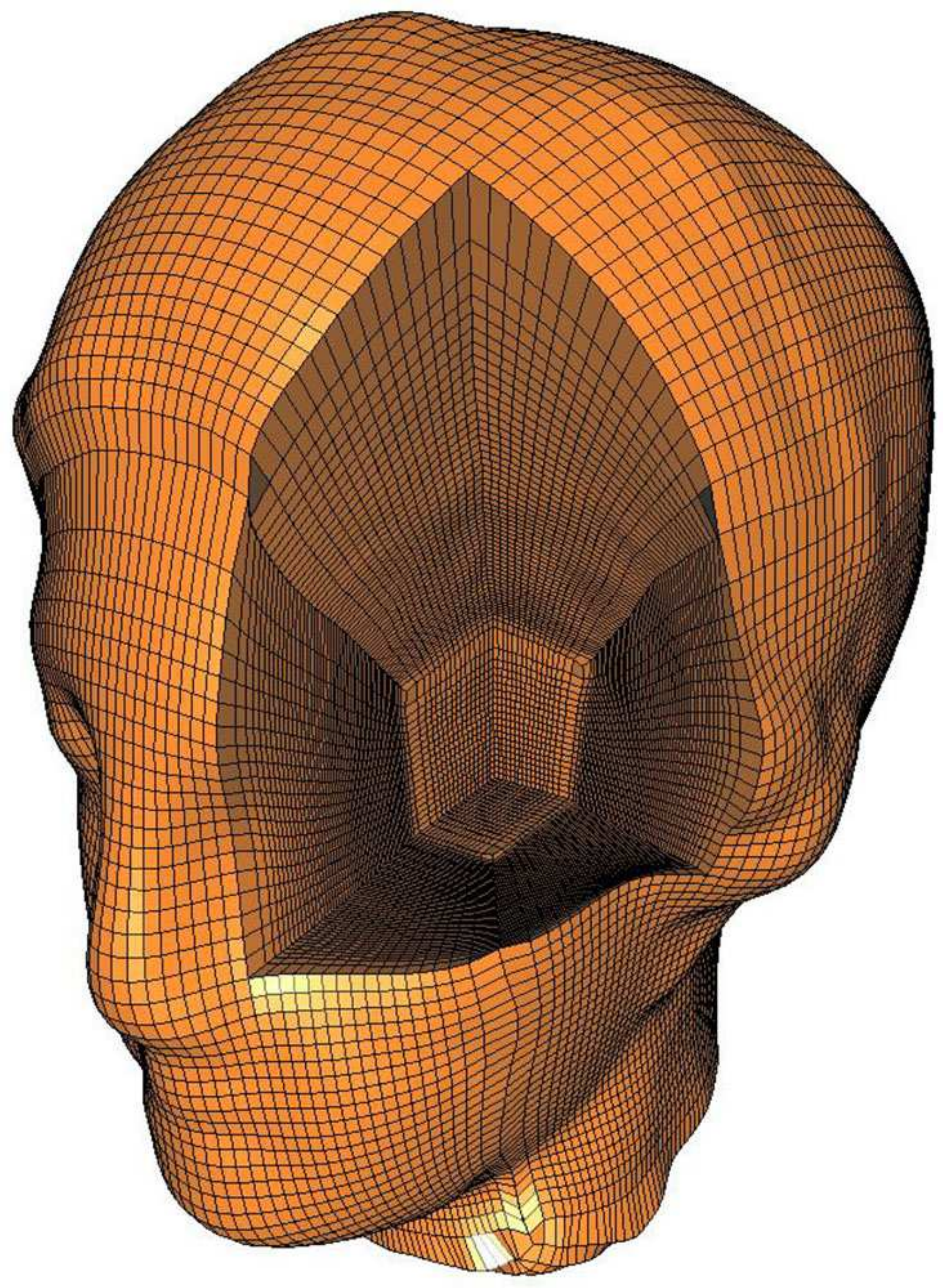}}
  \caption{The cut-away views of TBSs generated by our method.
  (a) Mannequin. (b) Tooth. (c) Duck. (d) Moai. (e) Ball joint.  (f) Venus.}
   \label{fig:hex}
\end{figure}
%------------------------------------------------------------------------------

 %----------------------------------------------------------------------------
 % Subsection:
 %----------------------------------------------------------------------------
 \section{Results}
 \label{sec:result}

\begin{figure}[!htb]
  \centering
  \subfigure[]{\label{subfig:mannequin_jac}
    \includegraphics[width=0.20\textwidth]{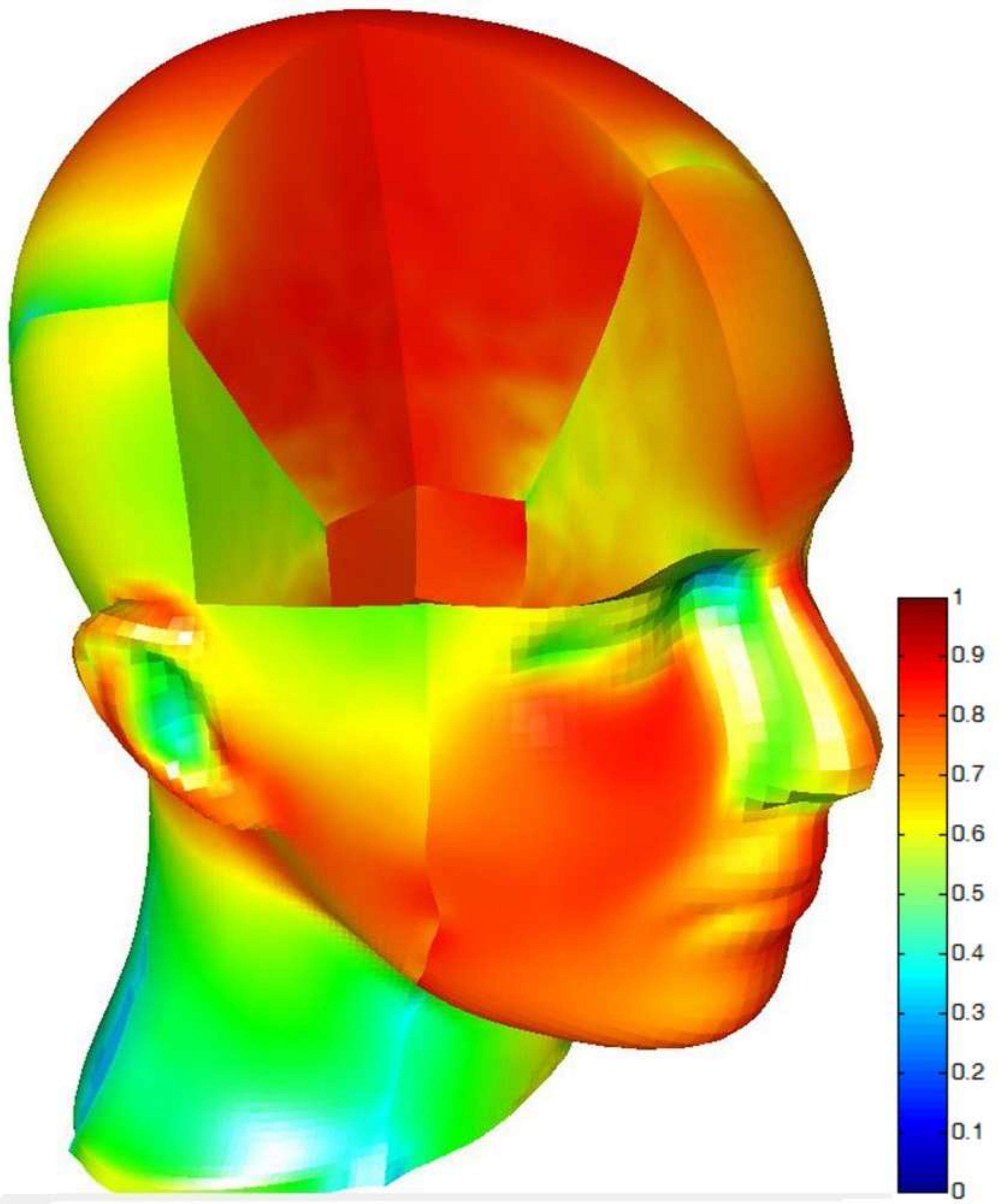}}
  \subfigure[]{\label{subfig:tooth_jac}
    \includegraphics[width=0.25\textwidth]{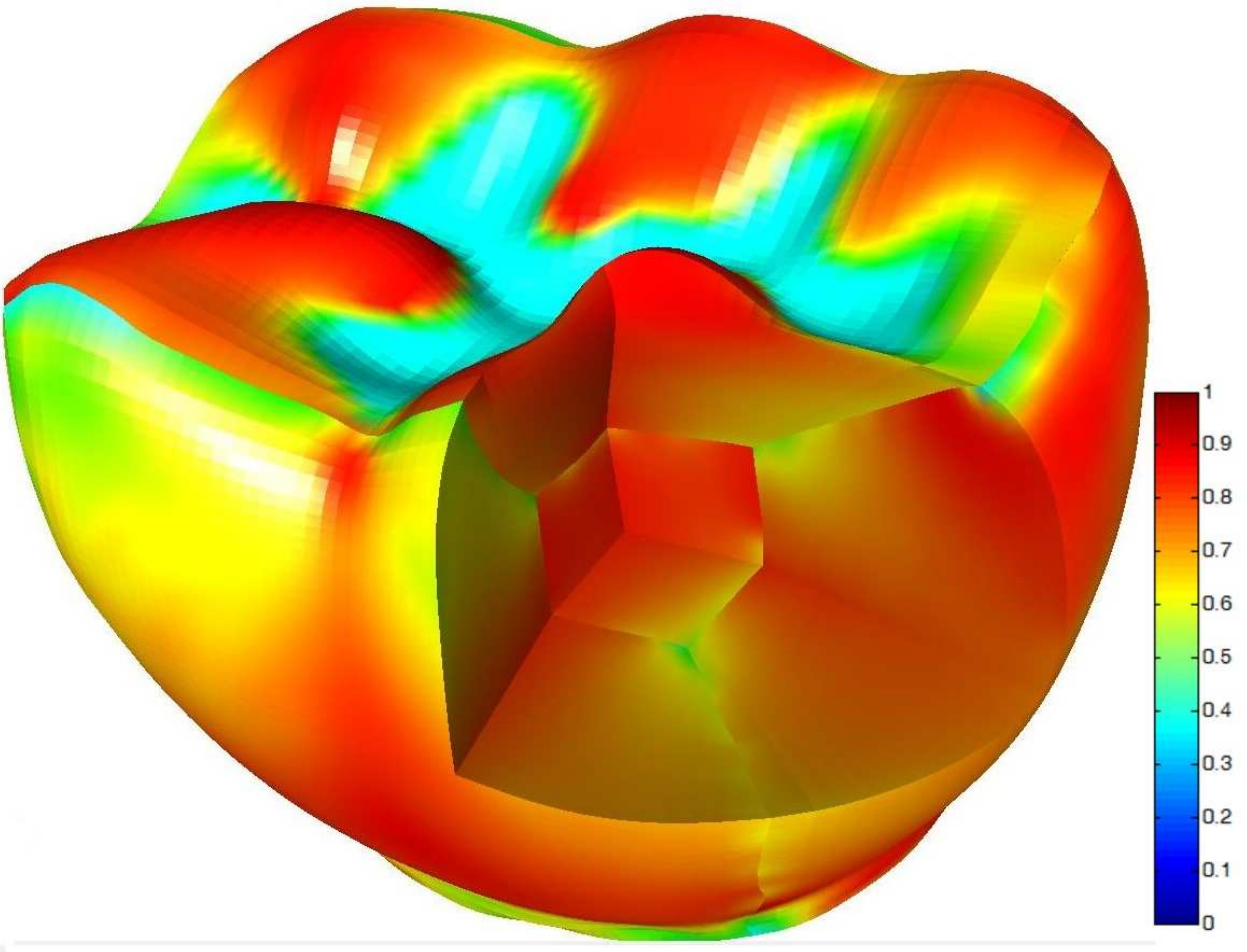}}
  \subfigure[]{\label{subfig:duck_jac}
    \includegraphics[width=0.24\textwidth]{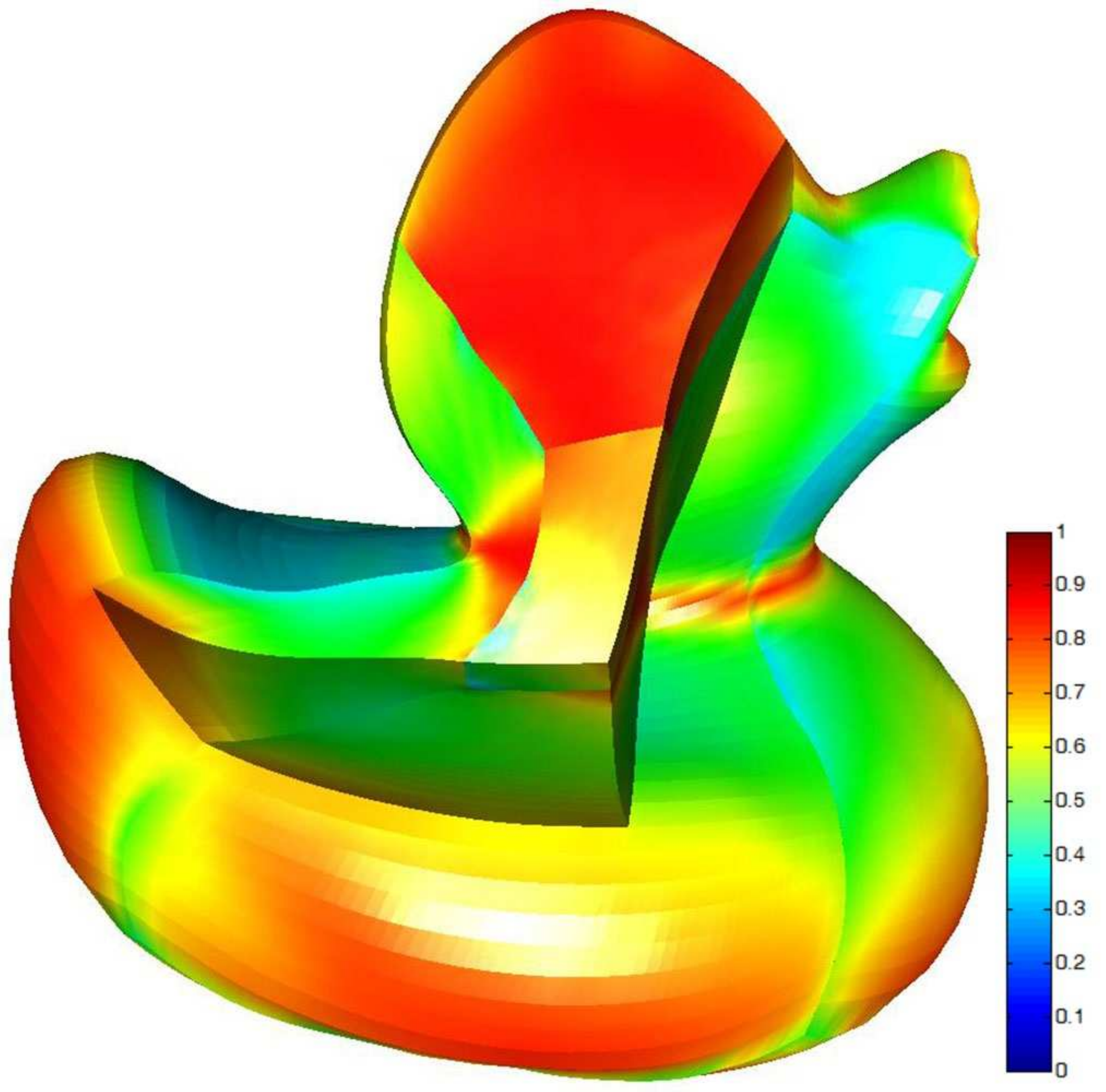}}
  \subfigure[]{\label{subfig:moai_jac}
    \includegraphics[width=0.17\textwidth]{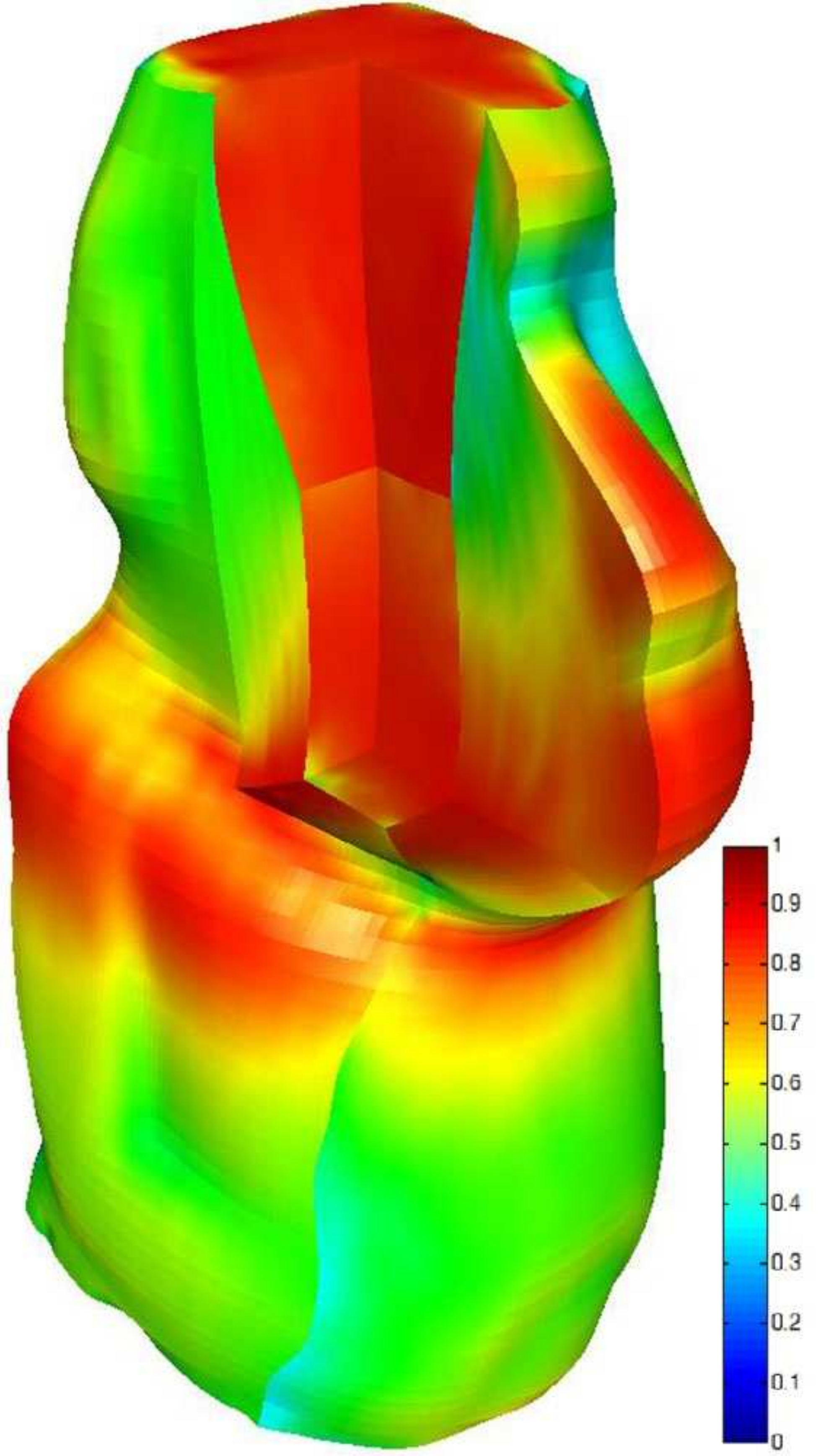}}
  \subfigure[]{\label{subfig:ball_joint_jac}
    \includegraphics[width=0.23\textwidth]{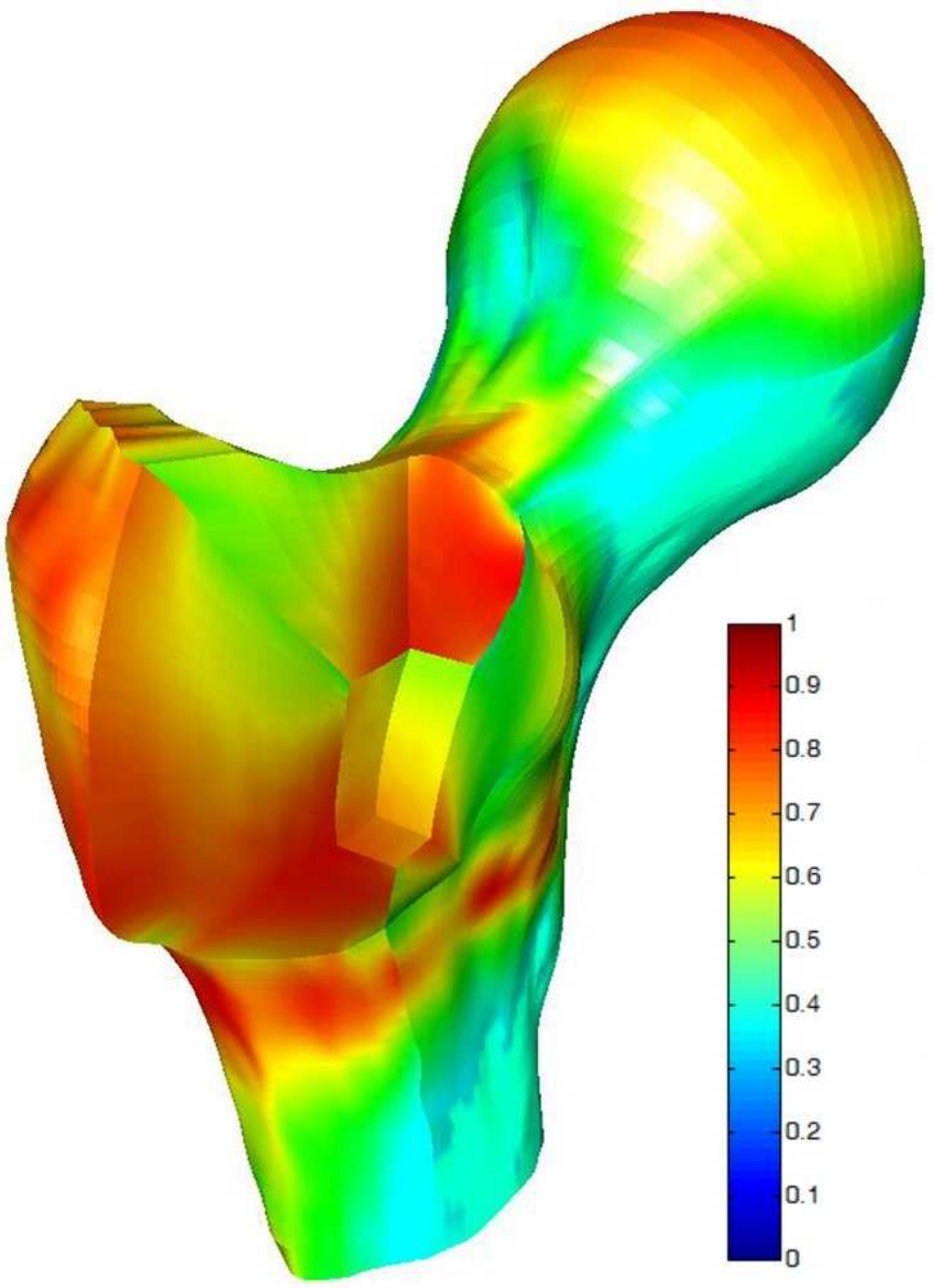}}
  \subfigure[]{\label{subfig:venus_jac}
    \includegraphics[width=0.21\textwidth]{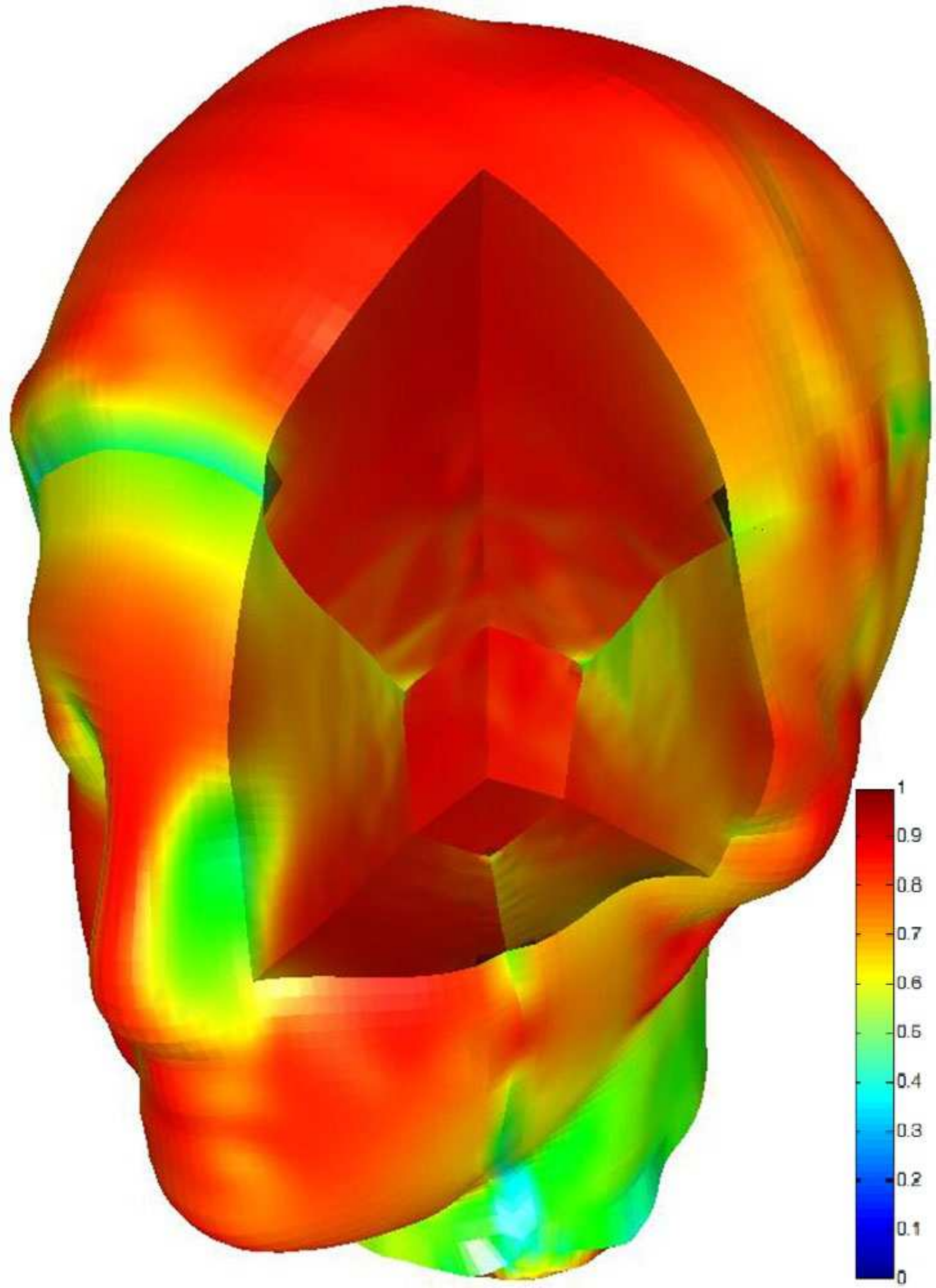}}
      \caption{The distribution of scaled Jacobian values on TBSs generated by our method.
    (a) Mannequin. (b) Tooth. (c) Duck. (d) Moai. (e) Ball joint. (f) Venus.}
   \label{fig:jac}
\end{figure}

 The algorithm developed in this paper is implemented using Visual Studio
    C++ 2010,
    and run on a PC with Intel Core2 Quad CPU Q9400 2.66 GHz and 4GB memory.
  In this paper, we demonstrated a method to generate six TBS models.
 In Fig.~\ref{fig:hex}, the cut-away views of the TBS models are illustrated.
 It can be seen that the iso-parametric curves vary smoothly not only inside
    a single TBS,
    but between two adjacent TBSs as well.
 Moreover, in Fig.~\ref{fig:jac},
    the distribution of the scaled Jacobian values~\cite{knupp2003method} of the TBSs is visualized with different colors.
 The darker the red color, the higher the scaled Jacobian values.
 As shown in Fig.~\ref{fig:jac}, the scaled Jacobian value of these models
    are all positive.
 Most region of the six TBS models is in red.
 Actually, in each of the six TBS models,
    the Jacobian values are larger than $0.5$ in over $80\%$ region.

%------------------------------------------------------------------------------
%-----------------------------Figure-------------------------------------------
%------------------------------------------------------------------------------

 The statistics of TBSs generated by our method are listed in
    Table~\ref{tbl:stat}.
 The third column of Table~\ref{tbl:stat} is the number of control points
    of TBSs, reading as $m \times n \times k \times l$.
 It means that, the number of control points of the central TBS is
    $m \times n \times k$,
    those of the six surrounding TBSs are $m \times n \times l$,
    $m \times n \times l$, $m \times k \times l$, $m \times k \times l$,
    $n \times k \times l$, and $n \times k \times l$, respectively.
 The fourth column is the fitting error,
    which is defined as,
    \begin{equation*}
      \frac{\sqrt{\sum_{l=0}^6 \frac{\sum_{i=0}^{N_i-1} \norm{\bm{H}_l(u_i,v_i,w_i)-\bm{V}_i}^2}{N_i}}}{L},
    \end{equation*}
    where $N_i$ is the number of tet mesh vertices in the $i^{th}$ sub-volume ($i=0,1,\cdots,6$),
    and $L$ is the diagonal length of the bounding box of the whole model.
 The fitting errors for all of the six models are in the order of magnitude
    $10^{-2}$.
 Moreover, in the fifth column,
     the ratios of the volume of the region with scaled Jacobian in $(0,0.2]$ to the whole volume of the TBS are listed.
 It can be seen that the largest ratio (for the model Ball joint)
    is $1.96\%$,
    and the ratios for the other models are all below $0.5\%$,
    meaning that regions with scaled Jacobian in $(0,0.2]$ in generated TBSs are very small.
 In the last four columns, we list the time (in second) cost in curve,
    surface, TBS fitting, and quality improvement (i.e., smoothness and fairness improvement).
 The whole process for the six models takes time ranging from 11 min to 37 min.

 Moreover, for comparison, the minimum scaled Jacobian values and
    average Jacobian values of TBSs generated by the method in Ref.~\cite{lin2015constructing} and our method, respectively,
    are presented in Table~\ref{tbl:comp}.
 The average Jacobian value is calculated as,
    \begin{equation*}
      avg Jac = \frac{\int \int \int_{\Omega} J(x,y,z) dx dy dz}
                     {\int \int \int_{\Omega} dx dy dz},
    \end{equation*}
    where $J(x,y,z)$ is the scaled Jacobian value~\cite{knupp2003method} at $(x,y,z)$.
 It can be seen from Table~\ref{tbl:comp} that,
    while the minimum scaled Jacobian values of TBS models generated by the
    method in~\cite{lin2015constructing} are all negative,
    those of TBS models produced by our method are all positive.
 In addition, with the two methods,
    the average scaled Jacobian values are comparable.

 %----------------------------------------------------------------------
 %------------------------Table-----------------------------------------
  \begin{table*}[!htb]
  \centering
  \footnotesize
  \caption{Statistical data of the TBSs generation method developed in this paper.}
  \label{tbl:stat}
  \begin{threeparttable}
  %\begin{center}
  \begin{tabular}{| c | c | c | c | c | c | c | c | c |}
  \hline
        \multirow{2}{*}{model} & \multirow{2}{*}{\#vert.\tnote{1}} & \multirow{2}{*}{\#control\tnote{2}} & \multirow{2}{*}{fitting error} & \multirow{2}{*}{ratio\tnote{3}} & \multicolumn{4}{|c|}{time\tnote{4}} \\
        \cline{6-9}
        &  &  &  & & curve  & surface   & TBS & quality \\
  \hline
  Mannequin & 44152 & $50 \times 50 \times 90 \times 30$ & $5.06 \times 10^{-2}$ & 0.11\% & 3.26 & 344.85 & 1268.87 & 605.72 \\
  \hline
    Tooth & 61311 & $30 \times 26 \times 26 \times 26$ &
    $5.07 \times 10^{-3}$ &  0.12\% & 1.62  & 198.54   & 464.13 & 124.66 \\
  \hline
    Duck  & 41998 & $30 \times 30 \times 24 \times 20$ &
    $7.65 \times 10^{-2}$ & 0.09\% & 1.37 & 166.14 & 393.67  & 117.90 \\
  \hline
    Moai & 8831 & $30 \times 30 \times 30 \times 20$ & $3.89 \times 10^{-2}$ & 0.34\% & 1.38 & 230.28 & 538.54 & 111.10 \\
  \hline
    Ball joint & 43994 & $30 \times 22 \times 34 \times 15$ & $2.37 \times 10^{-2}$ & 1.96\% & 1.41 & 280.62 & 685.98 & 130.53 \\
  \hline
    Venus & 35858 & $30 \times 30 \times 30 \times 24$ & $1.91 \times 10^{-2}$ & 0.00\% & 1.34 & 228.32 & 573.67 & 165.90 \\
  \hline
  \end{tabular}
 %\end{center}
 \begin{tablenotes}
    \item[1] Number of vertices of the input tet mesh model.
    \item[2] Number of the control points of the TBSs generated by our method.
    \item[3] The ratio of the volume of the region with scaled Jacobian in (0,0.2] to the whole volume of the TBS.
    \item[4] Time (in second) cost in curve, surface, TBS fitting and quality improvement (smoothness and fairness improvement).
 \end{tablenotes}
 %\end{center}
 \end{threeparttable}
 \end{table*}
 %%----------------------------------------------------------------------------

%\hline
%    Venus & 35858 & $30 \times 30 \times 30 \times 24$ &
%    $1.91 \times 10^{-2}$ &  &  &   &   & \\

 %%%%%%%%%%%%%%%%%%%%%%%%%%%%%%%%%%%%%%%%%%%%%%%%%%%%%%%%%%%%%%%%%%%%%%%%%%%%%%

 %----------------------------------------------------------------------
 %------------------------Table-----------------------------------------
  \begin{table}[!htb]
  \centering
  \footnotesize
  \caption{Comparison of the TBSs generated by our method and the method in Ref.~\cite{lin2015constructing}.}
  \label{tbl:comp}
  \begin{threeparttable}
  %\begin{center}
  \begin{tabular}{| c | c | c | c | c |}
  \hline
        \multirow{2}{*}{model} &
        \multicolumn{2}{|c|}{min Jac.\tnote{1}} &
        \multicolumn{2}{c|}{avg Jac\tnote{2}}\\
        \cline{2-5}
        &  our method & method in~\cite{lin2015constructing} & our method   &  method in~\cite{lin2015constructing} \\
  \hline
    Mannequin & 0.1048 & -0.9771 & 0.8713 & 0.8267 \\
  \hline
    Tooth & 0.1576  & -0.1171  & 0.9357 & 0.9321 \\
  \hline
    Duck  & 0.0851 & -0.9588 & 0.8154 & 0.8377 \\
  \hline
    Moai & 0.1127 & -0.9850 & 0.8412 & 0.8804 \\
  \hline
    Ball joint & 0.0987 & -0.5972 & 0.8367 & 0.8189 \\
  \hline
    Venus & 0.1243 & -0.7239  & 0.9176  & 0.8821 \\
  \hline
  \end{tabular}
 %\end{center}
 \begin{tablenotes}
    \item[1] Minimum scaled Jacobian value.
    \item[2] Average scaled Jacobian value.
 \end{tablenotes}
 %\end{center}
 \end{threeparttable}
 \end{table}

 It should be pointed out that, because the values of energy functions,
    i.e., $E_{fit}^c$ and $E_{val}^c$ in Eq.~\pref{eq:op_curve},
    $E_{fit}^s, E_u^s, E_v^s$, and $E_{uv}^s$ in Eq.~\pref{eq:op_surface},
    $E_{fit}^h, E_u^h, E_v^h, E_w^h, E_{uv}^h, E_{uw}^h$, and $E_{vw}^h$ in Eq.~\pref{eq:op_tbs},
    and $E_{smooth}^h, E_{fair}^h$ in Eq.~\pref{eq:op_smooth},
    differ much for different models,
    we choose different weights in solving the minimization problems (Eqs.~\pref{eq:op_curve}~\pref{eq:op_surface}~\pref{eq:op_tbs}~\pref{eq:op_smooth}),
    in order to balance the energy functions well,
    and achieve desirable results.
  The weights employed in generating the six TBSs are presented in
    Table~\ref{tbl:weight}.

%----------------------------------------------------------------------
 %------------------------Table-----------------------------------------
  \begin{table}[!htb]
  \centering
  \footnotesize
  \caption{Weights employed in generating TBSs.}
  \label{tbl:weight}
  \begin{threeparttable}
  %\begin{center}
  \begin{tabular}{| c | c | c | c | c | c | c | }
  \hline
              & Mannequin & Tooth & Duck & Moai & Ball joint & Venus \\
  \hline
  $\lambda_c$ & 0.0001 & 0.99 & 0.98 & 0.1 & 0.0001 & 0.0001 \\
  \hline
  $\lambda_s, \lambda_h$ & 0.0001 & 0.9 & 0.9 & 0.1 & 0.0001 & 0.0001 \\
  \hline
  $\mu_s, \mu_h$ & 0.0001 & 0.09 & 0.09 & 0.1 & 0.0001 & 0.0001 \\
  \hline
  $\lambda_f$ & 0.9 & 0.0001 & 0.0001 & 0.6 & 0.9 & 0.9 \\
  \hline
  \end{tabular}
 %\end{center}
 %\begin{tablenotes}
%    \item[1] Number of vertices of the input tet mesh model
% \end{tablenotes}
 %\end{center}
 \end{threeparttable}
 \end{table}
 %%----------------------------------------------------------------------

\begin{figure}[!htb]
  \centering
  \subfigure[]{\label{subfig:isis_hex}
    \includegraphics[width=0.11\textwidth]{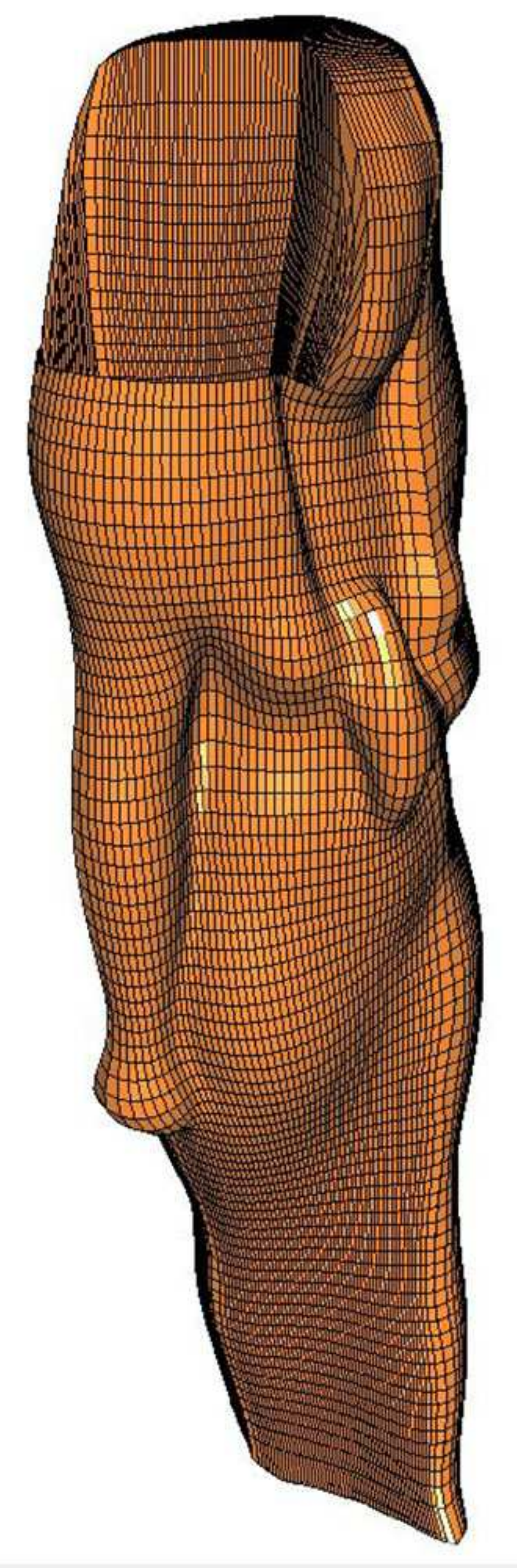}}
  \hspace{0.5cm}
  \subfigure[]{\label{subfig:isis_jac}
    \includegraphics[width=0.12\textwidth]{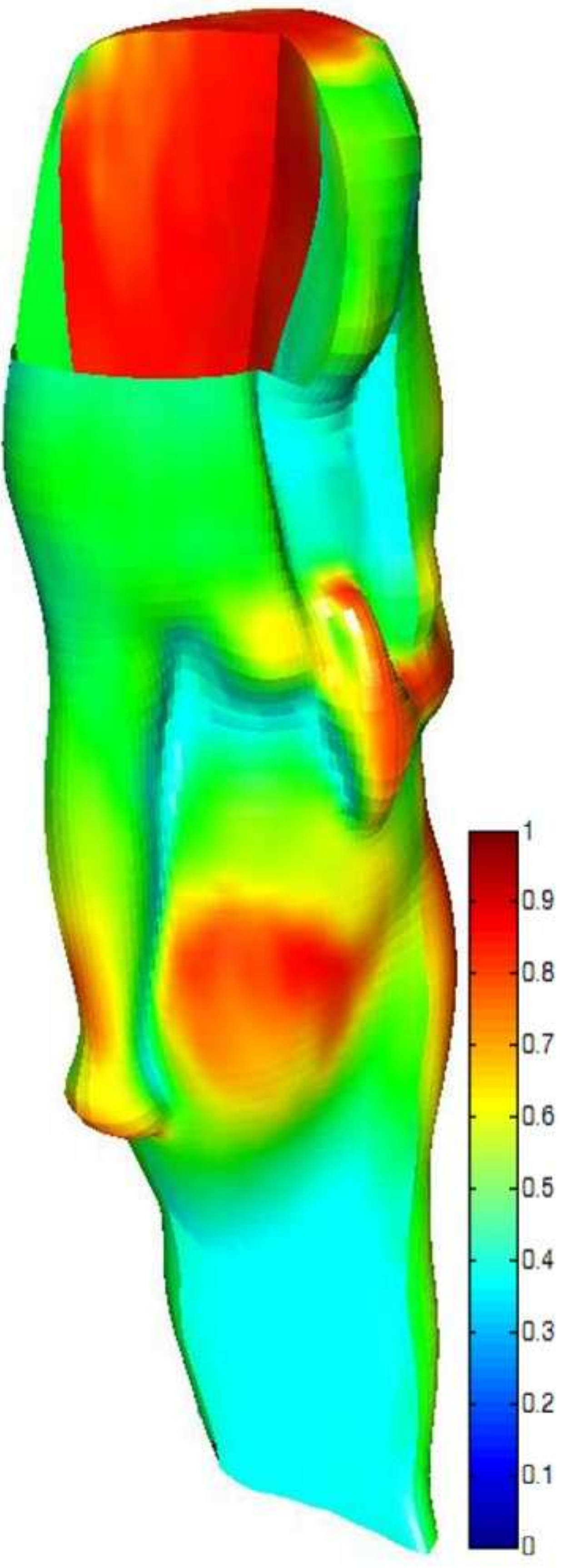}}
    \caption{The pillow operation is not suitable for some type of tet mesh model, and the Jacobian values in $16.69\%$ region of the \emph{Isis} TBS model lie in the interval $(0, 0.2]$. (a) Cut away view of the TBS model. (b) Distribution of the scaled Jacobian values.}
   \label{fig:isis}
\end{figure}

 \textbf{Limitations:} One limitation of the developed method lies that,
    the segmentation manner, i.e., the pillow operation,
    is not desirable for any tet mesh models.
 For example, because the bottom of the \emph{Isis} model shown in
    (Fig.~\ref{fig:isis}) is slim,
    the quality of the TBS generated by our method is not very good,
    though the Jacobian value can be guaranteed positive.
 The fitting precision can just reach the order of magnitude $10^{-1}$,
    and the Jacobian values in $16.69\%$ region lie in the interval $(0, 0.2]$.
 Therefore, more reasonable tet mesh segmentation method should be developed
    in the future.

 %------------------------------------------------------------------------------
%-----------------------------Figure-------------------------------------------
%------------------------------------------------------------------------------

 %----------------------------------------------------------------------------
 % Subsection:
 %----------------------------------------------------------------------------
 \section{Conclusion}
 \label{sec:conclusion}

 In this paper, we developed a method to generate a TBS by fitting a tet
    mesh model.
 The input to this method is a tet mesh model with six surfaces segmented on
    its boundary mesh.
 To improve the Jacobian values in the regions close to the boundaries,
    the tet mesh model was first partitioned into seven sub-volumes using the pillow operation.
 Then, a geometric iterative fitting algorithm was developed to fit the
    boundary curves, boundary surfaces, and the sub-volumes, separately.
 Because the validity conditions are integrated into the geometric
    iterations,
    the Jacobian values of the fitting boundary curves, boundary surfaces,
    and TBSs are guaranteed to be positive.
 Therefore, the generated TBSs are valid for isogeometric analysis.

\section*{Acknowledgement} This work is supported by the Natural Science Foundation of China
    (Nos. 61379072).

%-------------------------------------------------------------------------------
% Section: Appendix
%-------------------------------------------------------------------------------
\section*{Appendix: Convergence analysis of the GFD algorithm}

 In this appendix, the convergence analysis of the GFD algorithm
    (Algorithm~\ref{alg:gfd}) will be presented.
 Let,
 \begin{align*}
   D &= \left\{(\bm{P}_0, \bm{P}_1, \cdots, \bm{P}_m)\ |\
        \text{the constraints in~\pref{eq:min_problem} are satisfied} \right\}, \\
   L &= \left\{(\bm{P}_0, \bm{P}_1, \cdots, \bm{P}_m)\ |
    \ E(\bm{P}_0, \cdots, \bm{P}_m) \leq
    E(\bm{P}_0^{(0)}, \cdots, \bm{P}_m^{(0)})\right\}.
 \end{align*}
 The analysis of the convergence of the GFD algorithm
    (Algorithm~\ref{alg:gfd}) depends mainly on the following lemma~\cite{yaxiang1997theory}.

%------------------------------------------------------------------------------
% Lemma
%------------------------------------------------------------------------------
 \begin{lem} \label{lem:convergence}
  Suppose $\nabla E(\bm{P}_0, \bm{P}_1, \cdots, \bm{P}_m)$~\pref{eq:min_problem} is uniformly continuous on the region
  $D \cap L$,
  and the angle $\theta_k$ between the feasible direction $\bm{D}^{(k)}$ generated by the GFD algorithm (Algorithm~\ref{alg:gfd}) and
  $-\nabla E^{(k)}$ satisfies,
  \begin{equation*}
    \theta_k \leq \frac{\pi}{2} - \mu, \quad \text{for some}\ \mu > 0.
  \end{equation*}
  Then, $\nabla E^{(k)} = 0$ for some $k$, or
  $E(\bm{P}_0^{(k)}, \bm{P}_1^{(k)}, \cdots, \bm{P}_m^{(k)}) \rightarrow -\infty, (k \rightarrow \infty)$,
  or $\nabla E^{(k)} \rightarrow 0, (k \rightarrow \infty)$.
 \end{lem}

 Then, the convergence theorem for the GFD algorithm
    (Algorithm~\ref{alg:gfd}) is followed.

%------------------------------------------------------------------------------
% Theorem
%------------------------------------------------------------------------------
 \begin{thm} \label{thm:convergence}
 If $\nabla E(\bm{P}_0, \bm{P}_1, \cdots, \bm{P}_m)$~\pref{eq:min_problem}
    is uniformly continuous on the region $D \cap L$,
    and the objective function
    $E(\bm{P}_0, \bm{P}_1, \cdots, \bm{P}_m)$~\pref{eq:min_problem} is bounded,
    the GFD algorithm (Algorithm~\ref{alg:gfd}) is convergent.
 \end{thm}
 \textbf{Proof:} Denote $\theta_k$ as the angle between the feasible
    direction $\bm{D}^{(k)}$ and $-\nabla E^{(k)}$.
 In the iteration of the GFD algorithm,
    the following inequality holds,
    $$-\frac{\bm{D}^{(k)} \cdot \nabla E^{(k)}}
        {\norm{\bm{D}^{(k)}}_2\norm{\nabla E^{(k)}}_2} > \delta_d.$$
 So, there exists some $\mu >0$, such that,
    $$\theta_k \leq \frac{\pi}{2} - \mu.$$
 Therefore, based on Lemma~\ref{lem:convergence},
    together with that the objective function
    $E(\bm{P}_0, \bm{P}_1, \cdots, \bm{P}_m)$~\pref{eq:min_problem} is bounded,
    the GFD algorithm (Algorithm~\ref{alg:gfd}) is convergent. $\Box$

% ----------------------------------------------------------------
%\bibliographystyle{plain}
%\bibliographystyle{unsrt}
%\bibliographystyle{abbrv}
%\bibliographystyle{alpha}
\bibliographystyle{elsart-num}
%\nocite{*}
\bibliography{isogeometric}

\begin{thebibliography}{10}
\expandafter\ifx\csname url\endcsname\relax
  \def\url#1{\texttt{#1}}\fi
\expandafter\ifx\csname urlprefix\endcsname\relax\def\urlprefix{URL }\fi

\bibitem{hughes2005isogeometric}
T.~Hughes, J.~Cottrell, Y.~Bazilevs, Isogeometric analysis: {CAD}, finite
  elements, {NURBS}, exact geometry and mesh refinement, Computer methods in
  applied mechanics and engineering 194~(39) (2005) 4135--4195.

\bibitem{mitchell1995pillowing}
S.~A. Mitchell, T.~J. Tautges, Pillowing doublets: refining a mesh to ensure
  that faces share at most one edge, in: 4th International Meshing Roundtable,
  Citeseer, 1995, pp. 231--240.

\bibitem{zhang2007patient}
Y.~Zhang, Y.~Bazilevs, S.~Goswami, C.~L. Bajaj, T.~J. Hughes, Patient-specific
  vascular {NURBS} modeling for isogeometric analysis of blood flow, Computer
  methods in applied mechanics and engineering 196~(29) (2007) 2943--2959.

\bibitem{martin2009volumetric}
T.~Martin, E.~Cohen, R.~M. Kirby, Volumetric parameterization and trivariate
  {B-spline} fitting using harmonic functions, Computer Aided Geometric Design
  26~(6) (2009) 648--664.

\bibitem{xu2013analysis}
G.~Xu, B.~Mourrain, R.~Duvigneau, A.~Galligo, Analysis-suitable volume
  parameterization of multi-block computational domain in isogeometric
  applications, Computer-Aided Design 45~(2) (2013) 395--404.

\bibitem{wang2014optimization}
X.~Wang, X.~Qian, An optimization approach for constructing trivariate
  {B-spline} solids, Computer-Aided Design 46 (2014) 179--191.

\bibitem{aigner2009swept}
M.~Aigner, C.~Heinrich, B.~J{\"u}ttler, E.~Pilgerstorfer, B.~Simeon, A.-V.
  Vuong, Swept volume parameterization for isogeometric analysis, in:
  E.~Hancock, R.~Martin, M.~Sabin (Eds.), Mathematics of Surfaces XIII, Vol.
  5654 of Lecture Notes in Computer Science, Springer Berlin Heidelberg, 2009,
  pp. 19--44.

\bibitem{sederberg2004t}
T.~W. Sederberg, D.~L. Cardon, G.~T. Finnigan, N.~S. North, J.~Zheng, T.~Lyche,
  T-spline simplification and local refinement, in: Acm transactions on
  graphics (tog), Vol.~23, ACM, 2004, pp. 276--283.

\bibitem{escobar2011new}
J.~Escobar, J.~Casc{\'o}n, E.~Rodr{\'\i}guez, R.~Montenegro, A new approach to
  solid modeling with trivariate {T-splines} based on mesh optimization,
  Computer Methods in Applied Mechanics and Engineering 200~(45) (2011)
  3210--3222.

\bibitem{zhang2012solid}
Y.~Zhang, W.~Wang, T.~J. Hughes, Solid {T-spline} construction from boundary
  representations for genus-zero geometry, Computer Methods in Applied
  Mechanics and Engineering 249 (2012) 185--197.

\bibitem{wang2013trivariate}
W.~Wang, Y.~Zhang, L.~Liu, T.~J. Hughes, Trivariate solid {T-spline}
  construction from boundary triangulations with arbitrary genus topology,
  Computer-Aided Design 45~(2) (2013) 351--360.

\bibitem{liu2014volumetric}
L.~Liu, Y.~Zhang, T.~J.~R. Hughes, M.~A. Scott, T.~W. Sederberg, Volumetric
  {T-spline} construction using boolean operations, Engineering with Computers
  30~(4) (2014) 425--439.
\newline\urlprefix\url{http://dx.doi.org/10.1007/s00366-013-0346-6}

\bibitem{zhang2013conformal}
Y.~Zhang, W.~Wang, T.~J.~R. Hughes, Conformal solid {T-spline} construction
  from boundary {T-spline} representations, Computational Mechanics 51~(6)
  (2013) 1051--1059.
\newline\urlprefix\url{http://dx.doi.org/10.1007/s00466-012-0787-6}

\bibitem{burkhart2010iso}
D.~Burkhart, B.~Hamann, G.~Umlauf, Iso-geometric finite element analysis based
  on {Catmull-Clark} subdivision solids, in: Computer Graphics Forum, Vol.~29,
  Wiley Online Library, 2010, pp. 1575--1584.

\bibitem{hua2004multiresolution}
J.~Hua, Y.~He, H.~Qin, Multiresolution heterogeneous solid modeling and
  visualization using trivariate simplex splines, in: Proceedings of the ninth
  ACM symposium on Solid modeling and applications, Eurographics Association,
  2004, pp. 47--58.

\bibitem{li2010generalized}
B.~Li, X.~Li, K.~Wang, H.~Qin, Generalized polycube trivariate splines, in:
  Shape Modeling International Conference (SMI), 2010, IEEE, 2010, pp.
  261--265.

\bibitem{wang2012restricted}
K.~Wang, X.~Li, B.~Li, H.~Xu, H.~Qin, Restricted trivariate polycube splines
  for volumetric data modeling, Visualization and Computer Graphics, IEEE
  Transactions on 18~(5) (2012) 703--716.

\bibitem{sitetgen}
H.~Si, Tetgen: A quality tetrahedral mesh generator and three-dimensional
  delaunay triangulator, 2007, URL http://tetgen. berlios. de.

\bibitem{schoberl1997netgen}
J.~Sch{\"o}berl, Netgen an advancing front 2d/3d-mesh generator based on
  abstract rules, Computing and visualization in science 1~(1) (1997) 41--52.

\bibitem{lin2015constructing}
H.~Lin, S.~Jin, Q.~Hu, Z.~Liu, Constructing {B-spline} solids from tetrahedral
  meshes for isogeometric analysis, Computer Aided Geometric Design 35 (2015)
  109--120.

\bibitem{lin2004constructing}
H.~Lin, G.~Wang, C.~Dong, Constructing iterative non-uniform {B-spline} curve
  and surface to fit data points, SCIENCE IN CHINA, Series F 47~(3) (2004)
  315--331.

\bibitem{lin2005totally}
H.~Lin, H.~Bao, G.~Wang, Totally positive bases and progressive iteration
  approximation, Computers and Mathematics with Applications 50~(3-4) (2005)
  575--586.

\bibitem{lin2011extended}
H.~Lin, Z.~Zhang, An extended iterative format for the progressive-iteration
  approximation, Computers \& Graphics 35~(5) (2011) 967--975.

\bibitem{deng2014progressive}
C.~Deng, H.~Lin, Progressive and iterative approximation for least squares
  {B-spline} curve and surface fitting, Computer-Aided Design 47 (2014) 32--44.

\bibitem{shi06iterative}
L.~Shi, R.~Wang, An iterative algorithm of {NURBS} interpolation and
  approximation, Journal of Mathematical Research and Exposition 26~(4) (2006)
  735--743.

\bibitem{lin2013efficient}
H.~Lin, Z.~Zhang, An efficient method for fitting large data sets using
  {T-splines}, SIAM Journal on Scientific Computing 35~(6) (2013) A3052--A3068.

\bibitem{cheng2009loop}
F.~Cheng, F.~Fan, S.~Lai, C.~Huang, J.~Wang, J.~Yong, Loop subdivision surface
  based progressive interpolation, Journal of Computer Science and Technology
  24~(1) (2009) 39--46.

\bibitem{fan2008subdivision}
F.~Fan, F.~Cheng, S.~Lai, Subdivision based interpolation with shape control,
  Computer Aided Design \& Applications 5~(1-4) (2008) 539--547.

\bibitem{chen2008progressive}
Z.~Chen, X.~Luo, L.~Tan, B.~Ye, J.~Chen, Progressive interpolation based on
  {Catmull-Clark} subdivision surfaces, Pacific Graphic 2008, Computer Grahics
  Forum 27~(7) (2008) 1823--1827.

\bibitem{kineri2012b}
Y.~Kineri, M.~Wang, H.~Lin, T.~Maekawa, B-spline surface fitting by iterative
  geometric interpolation/approximation algorithms, Computer-Aided Design
  44~(7) (2012) 697--708.

\bibitem{yoshihara2012topologically}
H.~Yoshihara, T.~Yoshii, T.~Shibutani, T.~Maekawa, Topologically robust
  {B-spline} surface reconstruction from point clouds using level set methods
  and iterative geometric fitting algorithms, Computer Aided Geometric Design
  29~(7) (2012) 422--434.

\bibitem{okaniwa2012uniform}
S.~Okaniwa, A.~Nasri, H.~Lin, A.~Abbas, Y.~Kineri, T.~Maekawa, Uniform
  {B-spline} curve interpolation with prescribed tangent and curvature vectors,
  IEEE transactions on visualization and computer graphics 18~(9) (2012)
  1474--1487.

\bibitem{lin2014affine}
H.~Lin, Y.~Qin, H.~Liao, Y.~Xiong, Affine arithmetic-based {B-Spline} surface
  intersection with gpu acceleration, IEEE transactions on visualization and
  computer graphics 20~(2) (2014) 172--181.

\bibitem{farin2002curves}
G.~E. Farin, Curves and surfaces for CAGD: {A} practical guide, Morgan
  Kaufmann, 2002.

\bibitem{yaxiang1997theory}
Y.~Yuan, W.~Sun, Theory and method of optimization, Beijing: Science Publishing
  House, 1997.

\bibitem{burden2001numerical}
R.~Burden, J.~Faires, Numerical analysis, Thomson Brooks/Cole, Boston, 2001.

\bibitem{knupp2003method}
P.~Knupp, A method for hexahedral mesh shape optimization, International
  journal for numerical methods in engineering 58~(2) (2003) 319--332.

\end{thebibliography}

\end{document}